\documentstyle[preprint,aps,eqsecnum]{revtex}
\begin{document}
\draft
\title{
Chern-Simons Theory of the Anisotropic Quantum Heisenberg
Antiferromagnet on a Square Lattice}
\author{Ana Lopez$^{a}$, A.~G.~Rojo$^{b,c}$ and Eduardo Fradkin$^{a}$}
\address{$^a$Department of Physics, University of Illinois at
Urbana-\-Champaign , 1110 West Green Street,Urbana, IL 61801-3080;
$^b$The James Franck Institute, University of Chicago, 5640 S.~Ellis
Ave.~, Chicago, IL 60637-1467 and $^c$Department of Physics, University
of Michigan Ann Arbor, MI 48109-1120}
\bigskip

\maketitle

\begin{abstract}
We consider the anisotropic quantum Heisenberg antiferromagnet
(with anisotropy $\lambda$)
on a square lattice using a Chern-Simons (or Wigner-Jordan) approach. We
show that the Average Field Approximation (AFA) yields a phase diagram
with two phases: a Ne{\`e}l state for $\lambda>\lambda_c$ and a flux
phase for $\lambda<\lambda_c$ separated by a second order
transition at $\lambda_c<1$. We show that this
phase diagram does not describe the $XY$ regime of the antiferromagnet.
Fluctuations around the
AFA induce relevant operators which yield the
correct phase diagram.
We find an equivalence between the antiferromagnet and a relativistic
field theory of two self-interacting Dirac fermions coupled to a
Chern-Simons gauge field. The field theory has a
phase diagram with the correct number of
Goldstone modes in each regime and a phase transition at a critical
coupling $\lambda^* > \lambda_c$. We identify this transition with the
isotropic Heisenberg point. It has a non-vanishing
Ne{\` e}l order parameter, which drops to zero discontinuously for
$\lambda<\lambda^*$.
\end{abstract}

\bigskip


\narrowtext


\def\lnp{\raise.15ex\hbox{$/$}\kern-.57em\hbox{$p$}}
\def\lnk{\raise.15ex\hbox{$/$}\kern-.57em\hbox{$k$}}
\def\slp{{\raise.15ex\hbox{$/$}\kern-.57em\hbox{$\partial$}}}
\def\slD{\raise.15ex\hbox{$/$}\kern-.57em\hbox{$D$}}
\def \epi{e ^{i\vec{\pi}\cdot {\bf x}}}
\def \uq{u_{\bf q}}
\def \vq{v_{\bf q}}
\def \uk{u_{\bf k}}
\def \dk{\delta_{\bf k}}
\def \dq{\delta_{\bf q}}
\def \vk{v_{\bf k}}
\def\de{$\Delta E$}
\def\wfn{$\psi$}
\def\conf{{\bf r_1,r_2,\ldots,r_N}}
\def\exch{{\bf r_2,r_1,\ldots,r_N}}
\def\bold#1{${\mib #1}$}
\def\td{two dimensions}
\def\tdal{two-dimensional}
\def\qp{quasiparticle}
\def\qps{quasiparticles}
\def\mft{mean-field theory}
\def\odlro{${\rm ODLRO}^*$}
\def\as{\alpha_s}
\def\tas{\widetilde{\as}}
\def\amf{\alpha_{mf}}
\def\sc{superconductivity}
\def\scer{superconductor}
\def\scing{superconducting}
\def\scers{superconductors}
\def\musr{$\mu$SR}
\def\mnot{M_0}
\def\mnn{M_{00}}
\def\mmnn{$\mnn$}
\def\mmn{$\mnot$}
\def\mo{M^{(1)}}
\def\mt{M^{(2)}}
\def\fs{fractional statistics}
\def\bc{boundary conditions}
\def\xn{{\bf x}}



\section{Introduction}
\label{sec-intro}

Since the discovery of high-\-$T_c$ superconductors\cite{bednorz}, the
two-\-dimensional quantum Heisenberg model has received considerable
attention.
This is largely due to well established experimental facts which
strongly suggest that these compounds can be described by a
doped Heisenberg spin-$1/2$ quantum antiferromagnets \cite{anderson}.

Dimensionality plays a crucial role in the properties of the
Quantum Heisenberg Antiferromagnet.
The $S=1/2$ quantum antiferromagnetic chain can be solved
exactly using the Bethe ansatz\cite{bethe}.
By using the Wigner-\-Jordan transformation\cite{jordan,liebmattis},
this model can be mapped onto a system of spinless, interacting,
fermions with a coupling constant equal to the anisotropy parameter.
This model is particularly simple in the $XY$ limit where the spin problem
maps to free fermions.
Although fairly reliable in general dimensions, in one space dimension
the spin wave theory is plagued by a number of notorious problems.
 This approximation is based on the
Holstein-\-Primakov\cite{holstein}
 transformation which maps $S=1/2 $ spins
into hard core bosons.  The spin-\-wave approximation\cite{kittel}
relaxes the hard core constraint  and treats correctly the commutation
relation between spins in different sites. For
one-\-dimensional systems, spin-\-wave theory (or rather, the
$1/S$ expansion) is infrared divergent order-by-order. This
divergence is the manifestation of the fact that the continuous
symmetry of global spin rotations cannot be broken in one space
dimension. It also misses the essential fact
that half-\-integer spin systems are critical while integer spin systems
are always quantum disordered and have an energy gap\cite{haldanegap}.
These properties of the exact ground state of the system can be
recovered in one-\-dimensional
spin systems by using non-perturbative methods, such as the
Wigner--Jordan
transformation combined with bosonization\cite{luther,duncan}.

Much less is known for two-dimensional quantum
antiferromagnets. Firstly, there is no exact solution available
in any limit of the spin-${\frac {1}{2}}$ system. Spin-wave theory
predicts
a Ne{\`e}l ordered ground state for the isotropic antiferromagnet on a
square lattice, although with a moment reduced to $50\%$ of the
classical value by quantum fluctuations\cite{spinwave}.
The Hamiltonian for the anisotropic quantum Heisenberg antiferromagnet
on a square lattice is
\begin{equation}
H=J\sum_{<{\bf x},{\bf x} '>}\left\{\lambda S_z({\bf x})S_z({\bf x} ') +{1
\over
 2}\left[S^+({\bf x}   )S^-({\bf x}
')+S^-({\bf x} ')  S^+({\bf x} ')
\right] \right\}\ \ ,
\label{eq:ham1}
\end{equation}
where $<{\bf x} >,{\bf x} '>$ denotes nearest neighboring sites
and $\lambda$ is the
anisotropy parameter ($\lambda=1$ corresponds to the isotropic case).

The following facts are known to be true for this system.
For $\lambda \gg
1$, the Ising term dominates and the ground state should be close to a
{\em classical} antiferromagnet which has total $S_z=0$. This state has
an energy
gap and an expansion in powers of $1/\lambda$ is rapidly convergent.
In the opposite $XY$ limit, where $\lambda \to 0$, there is a
theorem\cite{xy}  due to Kennedy, Lieb and Shastry and to Kubo and
Kishi, which proves that there exists long range order with the spins
lined up {\em on the XY plane} for $\lambda < \lambda_1$ (with
$\lambda_1\geq 0.13$). The same theorem proves that, in the Ising
regime, the antiferromagnetic ground state extends {\em at least} down
to an anisotropy parameter $\lambda \leq 1.78$. No theorem is known
for the isotropic case $\lambda = 1$ and spin $S={\frac {1}{2}}$. For $S \geq
1$ Dyson, Lieb and Simon\cite{dyson} proved a theorem which shows that
there is Ne{\`e}l order {\em even} at the isotropic antiferromagnetic
point.
Finite size diagonalization\cite{reger}, quantum Monte
Carlo\cite{manousakis} and variational estimates\cite{vmc},
are more consistent with a Ne{\`e}l antiferromagnetic ground state
for the isotropic antiferromagnet.
For two-\-dimensional quantum antiferromagnets, the semiclassical $1/S$
expansion is free of the infrared divergencies found in one
dimension.
This approach predicts that the low energy limit of the isotropic
antiferromagnet is a non-linear sigma model {\em without} a topological
term\cite{1/s}. This latter results have been confirmed by detailed
renormalization group studies\cite{chn} which yield an excellent
agreement with experiments on $La_2 Cu O_4$.

These results suggest that the anisotropic quantum
antiferromagnet has a phase diagram with just two phases:
(a)   a Ne{\`e}l state with Ising anisotropy for $\lambda>1$, and (b) an
$XY$ phase for $\lambda <1$. For $\lambda>1$ the Ising anisotropy
should make all excitations massive ({\it i.e.}\ no Goldstone bosons for
$\lambda>1$). For $\lambda<1$ the $U(1)$ $XY$ symmetry is
spontaneously broken and there should be {\em one} Goldstone boson
(spin wave). In this scenario, exactly {\em at} the isotropic point
$\lambda=1$, the $SU(2)/U(1)$ global symmetry of the Heisenberg model
is spontaneously broken and there should be {\em two} Goldstone bosons
(spin waves), as predicted by the non-linear sigma model\cite{1/s}. In
some sense there is a phase transition at $\lambda=1$ in that the
Ne{\`e}l order parameter should jump {\em discontinuously} to zero as
$\lambda$ is {\em decreased} through $\lambda=1$. Precisely {\em at}
$\lambda=1$ the $XY$ and Ne{\`e}l orders are equivalent under an $SU(2)$
rotation and, thus, there is still long range order. In contrast, the
one-\-dimensional spin-${\frac {1}{2}}$ chain has a {\em line of fixed points}
for
$\lambda\leq 1$ and Ne{\`e}l order in the massive phase $\lambda>1$.
In section \ref{sec-1/s} we give an argument, based on the $1/S$
expansion, in support of this general scenario.

In this paper we investigate the anisotropic quantum Heisenberg
antiferromagnet on a square lattice using a generalized Wigner-\-Jordan
transformation constructed earlier by one of us\cite{fradkin1}.
In reference \cite{fradkin1} it was shown that the
Wigner-Jordan transformation in two dimensions is a special form of a
statistics changing transformation which, quite generally, is achieved
by coupling particles to (lattice) Chern-Simons gauge fields with a
properly chosen coupling constant $\theta$. The quantum
Heisenberg
antiferromagnet becomes equivalent to a system of spinless fermions
which interact with each other (just as in the case of the spin chain)
but which are also coupled to the Chern-Simons gauge field. Thus, even
in the $XY$ limit this system is interacting. Systems of this type
have been considered
recently in connection with the problem of anyon
superfluidity~\cite{laughlin,cwwh,anyons} and the Fractional Quantum Hall
Effect
(FQHE)~\cite{zhk,lopez}. Unlike the case of the spin chain, the
equivalent fermion problem is never free and a new type of approximation
has to be found. The analog of mean field theory in this context is
known as the Average Field Approximation (AFA) (defined below in
Section~\ref{sec-cslattice}). We will also show that the
physics of this problem is hidden  at the mean-field level and it is
only revealed by a careful consideration of the fluctuations.

Just as in systems of anyons or in the FQHE,
the energy
spectrum predicted by the AFA consists of free
fermions with an effective band
structure generated by the self consistent flux. We find that, for the
sector with $S_z=0$ (which corresponds to half-filling of either the
hard-core bosons or the fermions) the
average uniform effective flux is equal to one-half of the flux quantum.
Thus, on average, we find a {\em flux phase} analogous to that of the
gauge theory approach to quantum
antiferromagnets of Baskaran and Anderson\cite{pwa}, Affleck and
Marston\cite{affleck} and Kotliar\cite{kotliar}.
The AFA also predicts that, for $\lambda \geq \lambda_{\rm
c}$ (where $\lambda_{\rm c}\approx 0.39$), a gap opens up in the energy
spectrum. In this regime the fermion density {\em and} the flux
acquire a modulation with wave vector $(\pi,\pi)$. The Wigner-\-Jordan
transformation maps the $z$-component of the spin $S_z(\vec x)$ at $\vec
x$ to the fermion
occupation number $n(\vec x)$ by $S_z(\vec x)={\frac {1}{2}}-n(\vec
x)$. Thus, we identify this regime with Ne{\`e}l antiferromagnetic
order. For $\lambda < \lambda_{\rm c}$ the AFA spectrum of fermions is
massless.
Recently,
Wang\cite{wang} has studied the Heisenberg model on a
square lattice using the Wigner-\-Jordan transformation of reference
\cite{fradkin1} combined with an
approximation similar to the average field approximation discussed in
Section~\ref{sec-meanfield}. The results of the average field
approximation (AFA) that we present here disagree
with Wang's,  mainly  because his form of the AFA is not fully
self-\-consistent.

This mean-field spectrum is incompatible with the scenario
proposed above, based on the semiclassical expansion. There, the
spectrum
of low lying states contains {\it only} integer spin fluctuations ({\it i.e.}\
spin flips). Some of these states may be massless ( as for $\lambda \leq
1$) or all massive (such as for $\lambda >1$) but they are all {\it
bosonic}. In particular, and in contrast with one-dimensional systems,
the $1/S$ expansion predicts the existence of long range
order.

This problem is solved by a careful consideration of the role of the
fluctuations. As expected, symmetry plays a crucial role
here. The main problem is that, for small values of $\lambda$,
the AFA fermion spectrum is gapless. For arbitrary values of the
Chern-Simons coupling $\theta$, two types of gaps, even
or odd under Time Reversal ($T$) or Parity ($P$) can be generated by
fluctuations. The Lagrangian at the level of the AFA is even under both
$P$ and $T$. Thus, we should expect that fluctuations will generate all
terms with low scaling dimension ({\it i.e.}\  operators which are either
relevant or marginal) which are compatible with the symmetries of
the full system. Notice that here we encounter the {\em opposite} of the
situation
usually found with spontaneously broken symmetries where the mean-field
theory has {\em less} symmetry than the full system.

The symmetry
analysis becomes more transparent in terms of an effective theory for
the low energy degrees of freedom. This effective continuum theory
for the two-dimensional quantum antiferromagnet on a square lattice
turns out to be a theory of two species of
relativistic Dirac fermions (moving at the Fermi velocity defined in
Section~\ref{sec-critical}) coupled
to a Chern-Simons gauge field and to the fluctuations of an effective
Ne{\`e}l order parameter field. This effective field theory is
derived in Section~\ref{sec-critical}.
This theory is a generalization to 2D
of the well known equivalence between the antiferromagnetic spin
{\em chain}
and a field theory of interacting relativistic fermions in $1+1$
dimensions, known as the Luttinger-Thirring model.

In this language, the fermion spectrum can be understood in terms of the
possible energy gaps, or {\em masses}, of the fermions and of
the
symmetry properties of the operators connected with these masses. Our
analysis
shows that the Ne{\`e}l order parameter acts like a mass operator which
does not
break $P$ or $T$ and which we will denote as $M_{\rm Neel}$. In the
Ne{\`e}l phase the two species
of fermions acquire a mass, but with relative {\em opposite} sign.
However, for general values of the coupling constant $\theta$,
the Chern-Simons term breaks both $T$ and $P$.
Thus, quantum fluctuations of the Chern-Simons field will
necessarily generate all terms which break the same symmetries.
A fermion mass term with the {\em same sign} for both species also
breaks both $P$ and $T$ and we find that it does get generated by
quantum fluctuations.
We will refer to this as the {\em induced} fermion mass, $M_{\rm ind}$.
The actual phase diagram follows from the competition of these two
mechanisms.

The generation of a parity breaking mass term changes radically the
long distance behavior of the system.
It is well known~\cite{jackiw} that fermions with masses which break
time reversal and parity induce
Chern-Simons terms in the action of the gauge
fields at length scales
long
compared with the correlation length of the fermions, {\it i.e.}\ the inverse
of the parity breaking fermion mass. Also, the
sign of the {\em induced} Chern-Simons term is equal to the sign of the
mass of the fermion. Hence, at length scales long compared with
$1/M_{\rm ind}$, we expect to see a finite renormalization of the
Chern-Simons coupling constant from its bare value $\theta$ to some
effective value. This effective value depends on the pattern of
symmetry breaking, {\it i.e.}\ on the relative signs of
the induced masses which, in turn, are determined by the bare coupling
$\theta$ itself and by the nature of the ground state. We find that the
sign of $M_{\rm ind}$
is such that the induced Chern-Simons coupling tends to {\em reduce}
the bare Chern-Simons coupling.
This renormalization can be viewed as a tendency
to screen the bare statistics. This is a manifestation
of a more general property which we may think of as a ``Lenz Law of
Statistics"~\cite{steve}.

Thus, we arrive to the following scenario.
At small $\lambda$, the fermions acquire a parity breaking
mass $M_{\rm ind}$ through the fluctuations of the gauge field. In
turn, at distances long compared with $1/M_{\rm ind}$,
a Chern-Simons term is induced with a coupling constant which tends to
cancel
the bare statistics. We will find in Section~\ref{sec-phase}
that for $\theta=1/2\pi$
the cancellation is complete and the gauge fields are actually gapless
at long distances! This scenario is reminiscent of anyon
superfluidity. We will identify the gapless
transverse gauge fluctuation ({\it i.e.}\ Laughlin's mode~\cite{laughlin}) with
the gapless {\em transverse} spin wave of the $XY$ regime.
Clearly
two fermion masses with different symmetry necessarily compete with each
other. We expect that when they become of comparable magnitude
$\vert M_{\rm ind}\vert \approx \vert M_{\rm Neel} \vert $ a phase transition
should occur. For this range of $\lambda$, one of the two species of
fermions has
a mass that is very small and, at least nominally, is going to vanish at
some critical value of the anisotropy $\lambda^*$. We will
identify this phase transition with the $SU(2)$-symmetric Heisenberg
point\cite{matthew}. This phase
transition occurs at a value of the anisotropy $ \lambda^*>
\lambda_{\rm c}$. This phase transition preempts
the na{\" \i}ve second order transition predicted by the AFA from taking
place. This mechanism seems to bear a close analogy with a
fluctuation-\-induced first order transition~\cite{hlm,cw}.

To summarize, the Chern-Simons (or Wigner-Jordan) approach to the spin
$S={\frac {1}{2}}$
anisotropic quantum Heisenberg antiferromagnet yields a phase diagram which
is in qualitative agreement with the predictions of the semiclassical $1/S$
expansion. The physically correct ({\sl P.C.\/}) phase diagram is not
found at the level of
the Average Field Approximation and, except in the regime of strong Ising
anisotropy, it is due almost entirely to fluctuation effects. In this paper we
show how this physical picture is realized in the context of the Chern-Simons
theory. This is a (necessary) first step before these methods could be applied
to more subtle problem such as frustrated systems.

This paper is organized as follows. In Section~\ref{sec-1/s} we
develop the
semiclassical $1/S$ theory of the anisotropic antiferromagnet. In
Section~\ref{sec-cslattice} we review the Wigner-Jordan construction and
its connection with the Chern-Simons theory on a square lattice. In
Section~\ref{sec-meanfield} we present the results in the Average Field
Approximation and discuss the role of gaussian fluctuations. In
Section~\ref{sec-critical} the effective field theory is derived. The
dynamical, non-perturbative, effects of fluctuations are discussed in
Section~\ref{sec-phase} where we give a justification of the phase
diagram
discussed in this (long!) introduction. Section~\ref{sec-conclusions} is
devoted to the conclusions.

\section{The Anisotropic Heisenberg Quantum Antiferromagnet for large $S$}
\label{sec-1/s}

In this Section we discuss the semiclassical $1/S$ theory of the anisotropic
antiferromagnet in a square lattice.

The easiest way to get a path integral quantization for a spin system is to
use coherent states. In this Section we will follow the methods described in
reference  \cite{fradkin2}.

The set of coherent states $\{ |{\vec n}>\}$, labelled by the unit
vector
${\vec n}$, is generated by a rotation of the highest weight vector
($|S,S>$)
of an irreducible representation of the group SU(2) of spin $s$ of the
form
\begin{equation}
|\vec n> = e^{i \theta ({\vec n}_{0} \times {\vec n}). {\vec S}}
              |S,S>
\label{eq:ene1}
\end{equation}
where ${\vec n}_{0}$ is a unit vector along the quantization axis, $\theta$
is the colatitude ( ${\vec n}.{\vec n}_{0} = \cos \theta$) and $S_i$
($i=1,2,3$)
are the three generators of SU(2) in the spin-$s$ representation.
The state $\vert {\vec n} \rangle$ can be expanded in a complete basis
of the spin-$s$
irreducible representation $\{ |S,M>\}$ where $M$ labels the eigenvalues
of $S_3$. The coefficients of the expansion are the representation
matrices $D^{(S)}({\vec n})_{M,S}$
\begin{equation}
|\vec n> = \sum_{M=-S}^{S} D^{(S)}({\vec n})_{M,S} |S,M>
\label{eq:ene2}
\end{equation}
The matrices $D^{(S)}$ do not form a group but satisfy
\begin{equation}
D^{(S)}({\vec n}_1) \; D^{(S)}({\vec n}_2) = D^{(S)}({\vec n}_3)\;
                    e^{i \Phi ({\vec n}_{1},{\vec n}_{2},{\vec n}_{3}) S_{3}}
\label{eq:des}
\end{equation}
where ${\vec n}_{1}$, ${\vec n}_{2}$, and ${\vec n}_{3}$ are three arbitrary
unit vectors and $\Phi ({\vec n}_{1},{\vec n}_{2},{\vec n}_{3})$ is the area
of the spherical triangle with vertices at
${\vec n}_{1}$, ${\vec n}_{2}$, and ${\vec n}_{3}$.
Other useful properties of the spin coherent states are: (a) the inner
product $<{\vec n}_{1}|{\vec n}_{2}>$ \begin{equation}
<{\vec n}_{1}|{\vec n}_{2}>=
         e^{i \Phi ({\vec n}_{1},{\vec n}_{2},{\vec n}_{0}) s} \;
         {\left( { {1+ {\vec n}_{1}.{\vec n}_{2}}\over 2} \right) }^{s},
\label{eq:prop1}
\end{equation}
(b) the diagonal matrix elements of the generators ${\vec S}$
\begin{equation}
<{\vec n}| {\vec S}|{\vec n}> = s {\vec n}
\label{eq:prop2}
\end{equation}
and (c) the resolution of the identity operator
\begin{equation}
 {\hat I} = \int d\mu({\vec n}) \; |{\vec n}><{\vec n}|
\label{eq:prop3}
\end{equation}
where we have used the integration measure
\begin{equation}
 d\mu({\vec n}) = \left( {2s+1 \over {4\pi}}\right) \; d^3 {\vec n}\;
                  \delta ({\vec n}^2 - 1).
\label{eq:prop4}
\end{equation}
Using these properties we can write an expression for the path-integral in this
coherent state representation. The zero temperature partition function reads
\begin{equation}
{\cal Z} =  \int  {\cal D} {\vec n}\; e^{i{\cal S}_{M} [{\vec n}]} \ \ ,
\label{eq:zeta}
\end{equation}
where the action for the many-spin system in real-time is given by
\begin{equation}
{\cal S}_M [{\vec n}] = s \sum_{\vec r} {\cal S}_{WZ} [{\vec n}({\vec r})]
- {\int }^{T}_{0} dx_0 J s^2 \sum_{<{\vec r},{\vec r}'>}
\left\{ {\vec n}_{\perp }({\vec r},x_0).{\vec n}_{\perp }({\vec r}\; ',x_0)
+\lambda {\vec n}_3({\vec r},x_0).{\vec n}_3({\vec r}\; ',x_0) \right\}
\label{eq:act1}
\end{equation}
To write this expression we have used the Hamiltonian for the anisotropic
quantum Heisenberg antiferromagnet on a square lattice given by
eq.~(\ref{eq:ham1}) (with $<{\vec r},{\vec r}\; '>$ denoting nearest
neighboring sites), and we have assumed periodic boundary conditions.
In Eq.~(\ref{eq:act1}) ${\vec n}_{\perp }$ is the projection of
the vector $\vec n$ onto the $12$ (or $xy$) plane.

The first term in eq.~(\ref{eq:act1}) is just the sum of the Wess-Zumino terms
(or {\em Berry phases})
for the individual spins. The contribution of each term to the action is
\begin{equation}
{\cal S}_{WZ} [{\vec n}] = {\int }^{1}_{0} d\tau \; {\int }^{\beta }_{0} dt \;
            {\vec n}(t,\tau) . (\partial_{t} {\vec n}(t,\tau)
            {\times} \partial_{\tau} {\vec n}(t,\tau))
\label{eq:wz}
\end{equation}
where $\beta = iT$. The effect of this term is to quantize the spin.

The effective action ${\cal S}_M [{\vec n}]$ scales like $s$, the spin
representation. Therefore, in the large spin limit, the path integral
eq.~(\ref{eq:zeta}) should be dominated by the stationary points of the
action. This is the semiclassical limit. Corrections to the large-$s$ limit
can be arranged in an expansion in powers of $1\over s$.
Since we expect to be close to a N\'eel state, we will stagger the
configuration
\begin{equation}
{\vec n}({\vec r}) \rightarrow (-1)^{x_1 + x_2} {\vec n}({\vec r})
\label{eq:stag}
\end{equation}
The Wess-Zumino terms are odd under the replacement of eq.~(\ref{eq:stag})
and thus get staggered. Up to an additive constant the action reads
\begin{eqnarray}
{\cal S}_{M}[{\vec n}] &=& s \sum_{\vec r} (-1)^{x_1 +x_2}
{\cal S}_{WZ} [{\vec n}({\vec r})] \nonumber\\
&&- {J s^2\over 2} \sum_{\vec r}\sum_{j=1,2}{\int }_{0}^{T} dx_0
\left\{ [{\vec n}_{\perp }({\vec r},x_0)-
         {\vec n}_{\perp }({\vec r}+{\hat e}_j,x_0)]^2+
        [{\vec n}_{\perp }({\vec r},x_0)-
         {\vec n}_{\perp }({\vec r}-{\hat e}_j,x_0)]^2
\right\} \nonumber\\
&&+ \lambda
\left\{ [{\vec n}_{3 }({\vec r},x_0)-
         {\vec n}_{3 }({\vec r}+{\hat e}_j,x_0)]^2
         +[{\vec n}_{3 }({\vec r},x_0)-
         {\vec n}_{3 }({\vec r}-{\hat e}_j,x_0)]^2
\right\} \nonumber\\
&&+ (1-\lambda) \left\{ 2[{\vec n}_3({\vec r},x_0)]^2 +
                       [{\vec n}_3({\vec r}+{\hat e}_j,x_0)]^2 +
                       [{\vec n}_3({\vec r}-{\hat e}_j,x_0)]^2
 \right\}
\label{eq:act2}
\end{eqnarray}
We split the staggered spin field $\vec n$ in the following way
\begin{equation}
{\vec n}({\vec r}) = {\vec m} ({\vec r}) +
                   (-1)^{x_1 +x_2} a_{0} {\vec l}({\vec r})
\label{eq:split}
\end{equation}
where ${\vec m} ({\vec r})$ is a slowly varying piece , the order parameter
field, and ${\vec l} ({\vec r})$ is a small rapidly varying part which roughly
represents the average spin.
The constraint ${\vec n}^2 =1$ and the requirement that the order parameter
field $\vec m$ should obey the same constraint, ${\vec m}^2 =1$, demand
that ${\vec m}.{\vec l}=0$. Using this property  we can write the lagrangian
density for this theory in the long wavelength limit as
\begin{eqnarray}
&&{\cal L}_{M}({\vec m},{\vec l}) = {s\over a_0} {\vec l}.
     ({\vec m}{\times} \partial_{0} {\vec m}) \nonumber\\
&& - J s^2 \left\{ \sum_{j=1,2}[ (\partial_{j}{\vec m}_{\perp })^2 +
                                 \lambda (\partial_{j}{\vec m}_{3 })^2]
              + 8 [( {\vec l}_{\perp })^2 + \lambda ({\vec l}_3)^2]
              + 4 (1- \lambda )[{ ({\vec m}_{3 })^2 \over {a_0^2}}
                                     +({\vec l}_3)^2] \right\}
\label{eq:lag1}
\end{eqnarray}
where $a_0$ is the lattice spacing.

In the long wavelength limit, the Wess-Zumino action can be written as a sum
of a topological term and the first term in eq.~(\ref{eq:lag1}). It has
been shown~(\cite{1/s}) that if we expect to have N\'eel order,
the topological term does not contribute to the action in two space
dimensions.
Notice that, in the one-dimensional case, this same procedure leads
to a sigma model {\it with} a topological term.

After integrating out the fast modes, i.e., the components
${\vec l}_{\perp}$ and ${\vec l}_{3}$ of $\vec l$, in the partition function,
the resulting lagrangian is
\begin{eqnarray}
{\cal L}_{M} ({\vec m}) &=&
     {1\over 2g} [ {1\over v_s} ( \partial_0 {\vec m}_{\perp})^2 -
              v_s \sum_{j=1,2}( \partial_j {\vec m}_{\perp})^2] \nonumber\\
 &&  + {1\over 2g} [ {{1+ \lambda}\over {2 v_s} } ( \partial_0 {\vec m}_{3})^2
          - \lambda  v_s \sum_{j=1,2}( \partial_j {\vec m}_{3})^2] \nonumber\\
 &&- {8 (1- \lambda)  \over {a_0^2}}{v_s \over {2g}} ({\vec m}_{3})^2
   + { (1- \lambda) \over 2}{1\over {2 g v_s}} ({\vec m}_{3})^2
     [  ( \partial_0 {\vec m}_{\perp})^2 - ( \partial_0 {\vec m}_{3})^2]
\label{eq:lag2}
\end{eqnarray}
where the coupling constant $g$ and the spin wave velocity $v_s$ are given by
\begin{equation}
  g =  {2\over s} a_0 (1 + \lambda)^ {1\over 2}
\label{eq:ge}
\end{equation}
\begin{equation}
  v_s  =  4 J s  a_0 (1 + \lambda)^ {1\over 2}
\label{eq:vs}
\end{equation}
We can Wick-rotate back to imaginary time (i.e. $x_3 = i x_0$),
and write the Euclidean Lagrangian density ${\cal L}_E$ as
\begin{eqnarray}
{\cal L}_{E}({\vec m}) &=&
     {1\over 2g} [ {1\over v_s} ( \partial_3 {\vec m}_{\perp})^2 +
              v_s \sum_{j=1,2}( \partial_j {\vec m}_{\perp})^2]
  + {1\over 2g} [ {{1+ \lambda}\over {2 v_s} } ( \partial_3 {\vec m}_{3})^2
      + \lambda  v_s \sum_{j=1,2} ( \partial_j {\vec m}_{3})^2] \nonumber\\
 &&+ {8 (1- \lambda)  \over {a_0^2}}{v_s \over {2g}} ({\vec m}_{3})^2
   + { (1- \lambda) \over 2}{1\over {2 g v_s}} ({\vec m}_{3})^2
     [  ( \partial_3 {\vec m}_{\perp})^2 - ( \partial_3 {\vec m}_{3})^2]
\label{eq:lagE}
\end{eqnarray}
By direct inspection of  eq.~(\ref{eq:lagE}) one can see that
the third term of this action is relevant in the long wavelength limit. The
physics of this term is the following. For
$\lambda <1$, the system will lower its energy
by making $m_3 \rightarrow 0$, i.e., the system is in the $XY$ limit.
On the other hand, if $\lambda >1$ the energy will be maximized when $m_3$
acquires its maximum possible value, i.e., $m_3 =1$ and the system is
in the N\'eel state with Ising anisotropy.

The phase diagram for the anisotropic quantum antiferromagnet suggested by
these results has only two phases. For $\lambda >1$ the system is in a N\'eel
state with Ising anisotropy, and all the excitations are massive. For
$\lambda <1$ the systems is in an $XY$ phase, the U(1) $XY$ symmetry is
spontaneously broken and there should be one Goldstone boson. Exactly at
the isotropic point ($\lambda =1)$ there should be two Goldstone bosons
as predicted by the non linear sigma model \cite{1/s}.

\section{Chern-Simons on a Lattice}
\label{sec-cslattice}

We  begin by reviewing the path integral picture of a
spin system on a two dimensional lattice in terms of fermions coupled to
Chern-Simons gauge fields, introduced in reference \cite{fradkin1}.

The Wigner-\-Jordan transformation is based on the identification of a
system of {\em hard core bosons} ({\it i.e.}\ spin flips) with an equivalent
system of {\em fermions} each
of them rigidly attached with solenoids that carry one-\-half of the
flux quantum. Mathematically, the equivalent system is a theory of
fermions coupled to a Chern-Simons gauge field on the square lattice.
Chern-Simons theories\cite{wilczek} have been used with great success in
the Fractional Quantum Hall Effect\cite{zhk,lopez} and in
anyon superfluidity\cite{laughlin,cwwh,anyons}. The presence of the
lattice introduces a number of subtleties not found in continuum
systems. The role of the Chern-Simons gauge fields is to enforce the
constraint
that attaches particles to fluxes locally, {\it and} a set of
commutation relations among the gauge fields compatible with these
constraints\cite{eliezer}. These two features are key ingredients for
the Wigner-Jordan transformation to work.

However, unlike the one-\-dimensional spin chain,
in two dimensions the equivalent system of fermions is
coupled to a gauge field which can have {\em local} flux. Hence
the fermions are always
interacting, even in the $XY$ limit, and approximations become
necessary.
Written in the fermion language, the system can then be described in
terms of a theory of fermions which interact with each other and with a
Chern-Simons gauge field.

In what follows we will use a path-integral
description.
The zero temperature partition function for this problem has the form
\begin{equation}
{\cal Z} =  \int {\cal D} \psi^* {\cal D} \psi {\cal D} {\cal A} _\mu\;
e^{iS} \ \ , \label{eq:pf1}
\end{equation}
where $ \psi({\vec x},t)$ is a fermi field ({\it i.e.}\  Grassmann variables in
the path integral) defined on the sites $\{ {\vec x} \}$ of the square
lattice and $ {\cal A}_\mu$ are the statistical or Chern-Simons gauge
fields. The space components ${\cal A}_j({\vec x},t)$ are defined on the
links of the lattice while the time component ${\cal A}_0({\vec x},t)$
is defined on the sites. The role of the gauge field is to
change the statistics. The action $S$ is given by
\begin{eqnarray}
S &=& \int dt\left\{ \sum_{\bf x} \psi^*({\bf x},t) [iD_0 +\mu]
\psi({\bf
x},t) - {J\over 2} \sum_{j=1,2} \left(\left[\psi^*({\bf x}+e_j,t)e^{-i
 {\cal A}_j({\bf x},t)}\psi({\bf x},t)
+ {\rm c.c.} \right] \right. \right. \nonumber \\
& &  \left. \left.+ 2 \lambda (|\psi({\bf x},t)|^2-{1\over
 2})(|\psi({\bf x}+e_j,t)|^2-{1\over
2}) \right) \right\} + \theta S_{\rm cs}\left({\cal A}\right)
\label{eq:accion}
\end{eqnarray}
where
$D_0=\partial_0 +i {\cal A}_0$ is the covariant time derivative, and the
spatial covariant derivative is in this case the gauge covariant lattice
difference implied by the hopping term. The action $S$ of
Eq.~(\ref{eq:accion}) describes self-interacting fermions which are
coupled to a fluctuating Chern-Simons gauge field. The self-interaction
is represented by the quartic term in fermions in the action and it
corresponds to the $S_z S_z$ Ising interaction of the Heisenberg model.
We will call this term $S_{\rm int}$.

The lattice Chern-Simons $S_{\rm cs}\left({\cal A}\right)$  action was
defined in
references \cite{fradkin1,eliezer}. Its explicit form will be given
below. The coupling constant $\theta$ is chosen to
be ${\theta}={\frac {1}{2\pi}}$ so that the statistics corresponds to bosons
(with hard cores). For general Chern-Simons coupling $\theta$
this system is equivalent to a system of interacting anyons with
statistical angle $\delta \equiv {\frac {1}{2 \theta}}$, on a square
lattice\cite{fradkin1}. Lattice anyons have been studied numerically by
Canright
{\it et.~al.}\ ~\cite{girvin} and analytically by Fradkin
{}~\cite{anyons}. We will see
in Section~\ref{sec-meanfield} that the problem at hand is an example of
the degenerate solution found in reference \cite{anyons}.

In the representation of the Heisenberg model in terms of fermions
coupled to gauge fields, with the action of eq.~(\ref{eq:accion}), the
natural mean field approximation
consists of detaching the fermions from
their local
fluxes and to replace this dynamical flux by a static average
background.
Unlike spin-\-wave theory, in this mean field approximation
the hard core
constraint is taken into account exactly. The phase factors present
in the hopping amplitudes of the equivalent system of fermions, whose
role is to enforce the original
bosonic commutation relations, are treated approximately. The {\it
Average Field Approximation} (AFA), as this mean field theory has come
to
be known, was first introduced by Laughlin in the context of his study
of the anyon gas\cite{laughlin} and subsequently used quite extensively in
the context of the FQHE by two of us\cite{lopez}. A peculiar feature
of this mean field theory is
that it breaks a number of space-time symmetries in a very explicit
manner. For example, in the anyon gas the ground state obtained at the
AFA level breaks Galilean invariance while the actual ground state does
not. The fluctuations around the AFA restore Galilean invariance.
Likewise, in the context of the FQHE Galilean (or rather, {\it
magnetic}) invariance is broken at the level of the AFA but it is also
restored by gaussian fluctuations\cite{resfqhe}.

As stated above, the role of the Chern-Simons gauge fields is to change
the statistics from (hard-core) bosons to fermions. Here we will follow
the approach of references\cite{fradkin1,eliezer}. The effect of
the Chern-Simons action is twofold: (a) a constraint on the allowed
states which are required to satisfy a relation between the local
particle density and the local statistical flux
and (b) a set of commutation relations for the gauge
fields\cite{fradkin1}. With the sign conventions of the action
of eq.~(\ref{eq:accion}), the constraint reads
\begin{equation}
\rho({\vec x},t)=\theta {\cal B}({\vec x},t)
\label{eq:gauss}
\end{equation}
with $\rho({\vec x},t)\equiv \psi^*({\vec x},t) \psi({\vec x},t)$.
This is a constraint on the allowed states in the Hilbert space and it
plays the same role as Gauss' Law  in Maxwell's electrodynamics. In
eq.~(\ref{eq:gauss}) the particles live on the sites of the square
lattice whereas the flux ${\cal B}({\vec x},t)$ is defined on the sites of
the dual of the square lattice , {\it i.e.}\ the center of the plaquette
``north-east" of the lattice site $\vec x$. The
Chern-Simons gauge fields are defined to be on the links of the square
lattice.

The lattice form of the Chern-Simons action $S_{\rm cs}\left(
{\cal A}\right)$ can be written as the sum of two terms
\begin{equation}
S_{\rm cs}\left({\cal A}\right)=S_{\rm cs}^{(1)}+S_{\rm cs}^{(2)}
\label{eq:S_cs}
\end{equation}
where $S_{\rm cs}^{(1)}$ and  $S_{\rm cs}^{(2)}$ are responsible for
enforcing the constraint and for the determination of the commutation
relations respectively. Here we will use the
form given by Eliezer and Semenoff\cite{eliezer}
\begin{eqnarray}
S_{\rm cs}^{(1)} &=& \int \; dt \sum_{\vec x} A_0 \left( {\vec x},
t \right) \epsilon^{ij} d_i A_j \left( {\vec x}, t \right)
\label{eq:cons} \\
S_{\rm cs}^{(2)} &=& -{\frac {1}{2}} \int \; dt \sum_{\vec x}
A_i \left( {\vec x}, t \right) K^{ij}  {\frac {\partial}{\partial t}}
A_j \left( {\vec x}, t \right)
\label{eq:comrel}
\end{eqnarray}
Here, we have used the forward difference operator $d_i$ which acts on
functions $f(\vec x)$ defined on the sites as $d_i f(\vec x)\equiv f(\vec
x+{\hat e}_i)-f(\vec x)$, where ${\hat e}_i$ is a unit vector pointing
towards the direction $i=1,2$ of the square lattice . Similarly the
backward difference operator
${\hat d}_i$ acts like ${\hat d}_i\equiv f(\vec x)- f(\vec x-{\hat e}_i)$.
The kernel $K^{ij}$ is found to be given by the
matrix\cite{eliezer}
\begin{eqnarray}
K^{ij}=-{\frac {1}{2}} \left(
\begin{array}{cc}
d_2+{\hat d}_2                 & -2-2d_1+2{\hat d}_2+{\hat d}_2 d_1 \\
2+2d_2-2 {\hat d}_1-{\hat d}_1 d_2 & -d_1-{\hat d}_1
\end{array}
\right)
\label{eq:Kij}
\end{eqnarray}

The quartic term in the fermion part of the action represents the $S_z
S_z $ interaction. The constraint of Eq.~(\ref{eq:gauss}) restricts the
space of configurations to those in which the fermion occupation number
at a site is equal to the flux at the plaquette north-east of the site
divided by $\theta$. Hence, it is legitimate to replace in the
action the fermion density by $B/\theta$. Therefore, the interaction
term in the action, $S_{\rm int}$, becomes only a function of the
configuration of the gauge fields
\begin{equation}
S_{\rm int}=- {\frac {1}{2}} \sum_{{\bf x},{\bf x'}}
(\theta {\cal B} ({\bf x},t) -{\frac {1}{2}})
V({\bf x}- {\bf x}') (\theta {\cal B} ({\bf x}',t) -{\frac {1}{2}})
\label{eq:S_int}
\end{equation}
with a pair potential $V({\bf x}- {\bf x}')$ given by
\begin{equation}
V({\bf x}- {\bf x}')=
\left\{ \begin{array}{ll}
J \lambda & \mbox{if ${\bf x}'={\bf x}\pm {\bf e}_j$ ($j=1,2$)}
\nonumber \\
0 & \mbox{otherwise}
\end{array}
\right.
\label{eq:pair}
\end{equation}
Notice that the interaction term is {\em bilinear} in the gauge fields
instead of a quartic functional of the Fermi fields. This result was
obtained before, in the context of the FQHE, in reference\cite{resfqhe}.
Alternatively, one could use a Hubbard-Stratonovich transformation and
arrive to the same result\cite{resfqhe}.

By putting all the terms together we arrive to the final form of the
action
\begin{equation}
S=S_{\rm F}(\psi, \psi^*, {\cal A}_\mu)+S_{\rm int}({\cal A}_\mu) +\theta
S_{\rm
cs}({\cal A}_\mu)
\label{eq:finalaccion}
\end{equation}
where $S_{\rm F}(\psi, \psi^*, {\cal A}_\mu)$ is the action for the fermions
coupled to the gauge field
\begin{equation}
S_{\rm F}=\int dt \sum_{\bf x}\left\{ \psi^*({\bf x},t) [iD_0 +\mu]
\psi({\bf x},t) - {J\over 2} \sum_{j=1,2} \left[\psi^*({\bf x}+e_j,t)e^{-i
 {\cal A}_j({\bf x},t) ,t)}\psi({\bf x},t)
+ {\rm c.c.} \right]  \right\}
\label{eq:S_F}
\end{equation}
$S_{\rm int}({\cal A}_\mu)$ and $S_{\rm cs}$ are defined in
Eqs.~(\ref{eq:S_int}) and (\ref{eq:S_cs}) respectively. From now on we
will use the action of eq.~(\ref{eq:finalaccion}).

\section{ Mean Field Theory and Semiclassical Expansion}
\label{sec-meanfield}

We will now proceed to derive a mean field theory in the usual fashion.
The fermionic part in the action (\ref{eq:finalaccion}), being
bilinear, can be integrated out yielding a fermion determinant. The
resulting effective action $S_{\rm eff} $ is given by
\begin{eqnarray}
S_{\rm eff}&=& -i {\rm tr} \log [iD_0 + \mu - h({\cal A})]
+\theta S_{CS}({\cal A}_\mu) \nonumber \\
& & - {\frac {1}{2}} \int dt \sum_{{\bf x},{\bf x'}}
(\theta {\cal B} ({\bf x},t) -{\frac {1}{2}})
V({\bf x}- {\bf x}') (\theta {\cal B} ({\bf x'},t) -{\frac {1}{2}})
\label{eq:seff}
\end{eqnarray}
where $ h(\cal A)$ is the kinetic part that we can write in operator
form as
\begin{equation}
 h({\cal A})={\frac {J}{2}} \sum_{{\bf x}}\sum_{j=1,2} \
|{\bf x},t\rangle e^{i {\cal A}_j({\bf x},t)}\langle {\bf x}+{\bf
e}_j,t| + {\rm h.c.}
\end{equation}

The semiclassical approximation is obtained by expanding around
stationary configurations of fields that minimize the
 effective
action. It is worthwhile to note that this effective action does not
contain any small parameter to control  this expansion. This is a
problem that was also found in the context of the anyon superfluid
as well as in the fermion Chern-Simons approach to the FQHE of
references \cite{lopez,resfqhe}. There it was found that if the AFA
had a spectrum which is fully gapped, the fluctuations restore the
correct spectrum at long wavelengths. We will see that, for the problem
at hand, the AFA yields a gapless spectrum at least for a range of
values of the anisotropy parameter $\lambda$. Thus, the validity of
the AFA, even qualitatively, is questionable for that range. Indeed
we will find
it necessary to go well beyond the AFA in order to obtain asymptotically
exact results.

The Average Field Approximation is realized by the solutions of the
saddle-point equations
\begin{equation}
{\frac{\delta S_{\rm eff} }{\delta {\cal A}_\mu({{\bf x},t})}}
\Bigr|_{\bar{\cal A}}=0
\label{eq:spa}
\end{equation}
As usual, the variations of the fermionic part $S_{\rm F}$ of the action
$S_{\rm eff}$ ({\it i.e.}\ the first term in eq.~(\ref{eq:seff})) with
respect to the
components of the gauge field ${\cal A}_\mu$ gives AFA expressions for the
charge density $ n({\bf x},t)$
\begin{equation}
\langle n({\bf x},t)\rangle=
\langle \psi^*({\bf x},t)  \psi({\bf x},t)  \rangle =
-{\frac{\delta S_{\rm F} }{\delta {\cal A}_0({{\bf x},t})}}
\label{eq:fcharge}
\end{equation}
and current density $j_k({\bf x},t)$
\begin{equation}
\langle j_k({\bf x},t)\rangle=
\langle {\frac {iJ}{2}}\left[\psi^*({\bf x},t) e^{i{{\cal A}}_k({\bf x},t) }
\psi({\bf x}+{\bf e}_k,t)-\psi^*({\bf x}+{\bf e}_k,t) e^{-i{{\cal A}}_k({\bf
x},t) } \psi({\bf x},t) \right] \rangle=
-{\frac{\delta S_{\rm F} }{\delta {\cal A}_k({{\bf x},t})}}
\label{eq:fcurrent}
\end{equation}
Within the AFA, we find that the average density and currents are given
by
\begin{eqnarray}
\langle n({\bf x},t)\rangle_{\rm AFA}
&=&-{\frac{\delta S_{\rm F} }{\delta {\cal A}_0({\bf x},t)}}
= -iS({\bf x},t;{\bf x},t)
\label{eq:density}
\\
\langle j_k({\bf x},t)\rangle_{\rm AFA}
&=&-{\frac {\delta S_{\rm F} }{\delta {\cal A}_j({\bf x},t)}}
= {\frac {J}{2}} \left(
S({\bf x}+{\bf e}_j,t;{\bf x},t) e^{i {\bar {\cal A}}_j({\bf x},t)}-
S({\bf x},t;{\bf x} {\bf e}_j,t) e^{-i {\bar {\cal A}} _j({\bf x},t)}
\right)
\label{eq:current}
\end{eqnarray}
where ${\bar D}_0=\partial_0+i{\bar {\cal A}}_0$ and ${\bar
{\cal A}}_\mu$ ($\mu=0,1,2$)
is the expectation values of the
components
of the gauge fields within the AFA. The function $S({\bf x},t;{\bf
x}',t')$,
which appears in (\ref{eq:density}) and (\ref{eq:current}), is the
Green function for the fermions moving in the background field
${\bar {\cal A}}_\mu $
which is the solution of the lattice differential equation
\begin{equation}
\left(i{\bar D}_0 + \mu - h({\bar {\cal A}} ) \right)S({\bf
x},t;{\bf x}',t')= \delta_{{\bf x},{\bf x}'} \delta(t-t')
\label{eq:defgfn}
\end{equation}
Below we give the explicit form of this Green function.

By varying $S_{\rm eff} $ with respect to ${\cal A}_0$ we recover
the constraint equation, now as a condition for the
value of the local density of the stationary configurations
\begin{equation}
\langle n({\bf x} )\rangle = \theta \langle {\cal B}({\bf x}) \rangle
\label{eq:aa}
\end{equation}
Likewise, by differentiating with respect to the spacial components
${\cal A}_k$, we find an equation for the fermion current
\begin{equation}
\langle j_k({\bf x} ) \rangle = \theta \epsilon^{kl} d_l
\langle {\cal A}_0({\bf x}) \rangle + {\frac {\theta}{2}}
 \left( K^{lk}-K^{kl} \right) \partial_0
\langle {\cal A}_l({\bf x}) \rangle
- \theta^2 \epsilon_{kl} \sum_{{\bf x}'}
V({\bf x}-{\bf x}') {\hat d}_l \langle {\cal B} ({\bf x}') \rangle
\label{eq:minim}
\end{equation}
In terms of the pair potential of Eq.~(\ref{eq:pair}), we can write the
expectation value of the current in the form
\begin{eqnarray}
\langle j_k({\bf x} )\rangle = \theta \epsilon^{kl} d_l
\langle {\cal A}_0({\bf x}) \rangle &+& {\frac {\theta}{2}}
\left( K^{lk}-K^{kl} \right) \partial_0
\langle {\cal A}_l({\bf x}) \rangle \nonumber \\
&-& \theta^2 J \lambda \epsilon_{kl} {\hat d}_l
\sum_{j=1,2} \left( \langle {\cal B} ({\bf x}+{\bf e}_j)
\rangle+\langle {\cal B} ({\bf x}-{\bf e}_j ) \rangle \right)
\label{eq:minim2}
\end{eqnarray}
In Eqs..~(\ref{eq:minim}) and (\ref{eq:minim2}) $K^{lk}$ is the
operator matrix defined in Eq.~(\ref{eq:Kij}).
Using this definition explicitly, the terms in Eq.~(\ref{eq:minim2})
which depend on $K^{lk}$ are given by
\begin{equation}
\theta \left( K^{lk}-K^{kl} \right) \partial_0 \langle {\cal A}_l({\bf
x}) \rangle
= -{\frac{\theta}{2}}    \epsilon^{kl} \left[ 4+2 (d_2-{\hat  d}_2)+2
(d_1-{\hat d}_1)+{\hat  d}_2 d_1+{\hat  d}_1 d_2 \right] \partial_0
\langle {\cal A}_l({\bf x}) \rangle
\label{eq:terms}
\end{equation}
In the continuum limit, the first two terms in Eq.~(\ref{eq:minim2})
are equal to the conventional Chern-Simons current, {\it i.e.}\ $\theta
\epsilon^{kl} \langle {\cal E}_l(\bf x) \rangle$.

The Saddle Point Equations (\ref{eq:aa}) and (\ref{eq:minim2}) have
many solutions. For a half-filled system ({\it i.e.}\ $S_z=0$), the solution
with lowest energy corresponds
to a stationary state with a modulated charge density and with zero
current. In the magnetic language
this is a state with a non zero N\`{e}el order parameter $\Delta $,
 such that
\begin{equation}
\langle {n}({\bf x} )\rangle={\frac {1}{2}} -\Delta e^{i\vec{ \pi}\cdot {\bf x}
}.
\label{eq:nx}
\end{equation}
Using the constraint of Eq.~(\ref{eq:aa}) we get
\begin{eqnarray}
\langle B({\bf x} )\rangle &=&{\frac {1}{\theta}} \langle {n}(\xn )\rangle
\nonumber \\
                 & \equiv & 2 \pi \langle {n}({\bf x}  )\rangle
\label{eq:avflux}
\end{eqnarray}
This state is time independent and it does not support any current,
local or global. For
a square lattice with $N \times N$ sites (with
$N$ even), periodic boundary conditions and $\theta=1/2\pi$, we can
satisfy this static constraint with the following choice of gauge fields
\begin{equation}
\begin{array}{lcr}
{\cal A}_0= {\tilde A}_0 \epi   & \;\;\;\; & {\cal A}_j({\bf x}) =
\delta_{j,2}
\left({\pi\over 2} + {\vec \pi} \cdot {\vec x}+\pi \Delta \epi \right)
\end{array}
\label{eq:gauge}
\end{equation}
This solution corresponds to a problem in which a fermion moves in the
presence of a modulated magnetic field (with an average
of {\em half flux quantum per plaquette}) and a periodic potential
$\cal A_0$ with the {\em same} modulation.
By solving these equations we
find a solution in which
\begin{equation}
{\tilde A}_0=4\lambda J \Delta
\label{eq:A_0}
\end{equation}
For ground states with a
modulation in the effective magnetic field, a periodic site potential
$\langle {\cal A}_0\rangle$ appears which is commensurate with the variation
of the field. The charge and the field vary in the same way, as
required by the constraint. Thus, everything is self-consistent. Let us
note in passing that for sectors with $S_z \not=0$
the average flux is not equal to one-half of the quantum. At the level
of the AFA this problem now becomes equivalent to a general Hofstadter
problem\cite{hofstadter}. This problem is known to a have a very complex
spectrum which we will not explore here. One of us\cite{anyons}
discussed a similar problem in his treatment of the lattice anyon gas.
In what follows we will only discuss the case $S_z=0$.

The Green's functions of the saddle point problem  are  obtained
by solving the lattice differential equation of Eq.~(\ref{eq:defgfn}).
In the gauge that we have chosen, ${\bar {\cal A}}_1=0$ and with the
configuration of gauge fields of Eq.~(\ref{eq:gauge}), the solution of
Eq.~(\ref{eq:defgfn})
is the Green function  of a one-particle system with the effective
Hamiltonian $H_{\rm MF}=h({\bar {\cal A}}_j)+\sum_{{\bf x}}{\bar
{\cal A}}_0({\bf x}) $, which takes the form
\begin{equation}
H_{MF}=\sum_{\bf x} \left\{4\lambda J \Delta e^{i\vec{\pi}\cdot{\bf
x}}\;\; \vert {\bf x}\rangle \langle {\bf x}\vert
+ \frac{J}{2}\;\; \left( \vert{\bf x}\rangle
\langle{\bf x}+e_1\vert + a({\bf x}) \vert {\bf x} \rangle
\langle{\bf x}+e_2 \vert + {\rm h.c.} \right) \right\}
\label{eq:hmf}
\end{equation}
with
\begin{equation}
a({\bf x})=e^{i{\bar {\cal A}}_2({\bf x})}=- \sin \pi \Delta + i\epi
\cos \pi \Delta \ \ .
\end{equation}
Note that $a({\bf x})$ is equal to $a=ie^{i\pi \Delta}$
when ${\bf x} $ belongs to the $A$ sublattice and to
$a^*=-ie^{-i\pi \Delta}$
when ${\bf x} $ belongs to the $B$ sublattice.
For $\Delta = 0$ we recover the flux-\-phase state of
(uniform) half flux quantum per plaquette\cite{affleck,kotliar}.

The Hamiltonian of Eq.~(\ref{eq:hmf}) can be diagonalized in Fourier
space. The charge density and the effective flux are periodic functions
which take different values on the sublattices $A$ and $B$. Thus, the
Fermion Green function is a matrix whose entries label the sublattice
dependence of its arguments. It has the form
\begin{equation}
S_{\alpha \beta}(x,x')=
\left[
\begin{array}{cc}
S_{\rm AA}(x,x') & S_{\rm AB}(x,x')\\
S_{\rm BA}(x,x') & S_{\rm BB}(x,x')
\end{array}
 \right]
\end{equation}
where $x \equiv  ({\bf x},t)$ and
 $\alpha, \beta= {\rm A} , {\rm B}$. We find
\begin{equation}
S_{\alpha \beta}(x,x')=
\int_{-\infty}^{+\infty} {\frac {d \omega}{2 \pi}}
\int_{|k_i|\leq {\frac{\pi}{2}}}{\frac {d {\bf k}}{\pi^2}}
{\frac{e^{i \omega (t-t')-i({\bf x}- {\bf x}')}}
{\omega^2-E^2({\bf k})+i\epsilon}}
 \left(
\begin{array}{cc}
\omega+4J \lambda \Delta & J ( \cos k_1+ a \cos k_2)\\
J ( \cos k_1+ a^* \cos k_2) & \omega-4J \lambda \Delta
\end{array}
\right)
\label{eq:green}
\end{equation}
Notice that the momentum integrals are done over the range $|k_i| \leq
{\frac {\pi}{2}}$ ($i=1,2$). The fermion dispersion $E({\bf k})$ is
given by
\begin{equation}
E({\bf k})=+J \sqrt{\cos ^2 k_1 + \cos ^2 k_2 -2 \sin
(\pi \Delta)
\cos k_1 \cos k_2 + (4\lambda \Delta)^2 }
\label{eq:E_k}
\end{equation}
Recently, Wang \cite{wang} obtained similar results but without
self-consistency between charge and flux modulations.

The dependence of  $E({\bf
k})$ in $\pi \Delta $ as given by eq.~(\ref{eq:E_k}) is a consequence of
the self-consistency.
Once the Green function is known, the parameter $\Delta$ can be
calculated by demanding that the saddle-point equations be satisfied.
Thus, the average density on one sublattice, say $A$, must be given by
\begin{eqnarray}
\langle n({\bf x},t)\rangle_{\rm A}&=& -iS_{\rm AA}({\bf x},t;{\bf x},t)
=-i
\int_{-\infty}^{+\infty} {\frac {d \omega}{2 \pi}}
\int_{|k_i|\leq {\frac{\pi}{2}}}{\frac {d {\bf k}}{\pi^2}}
{\frac{ \omega+4J \lambda \Delta }
{\omega^2-E^2({\bf k})+i\epsilon}}
\nonumber \\
&=&{\frac {1}{2}}- 2J \lambda \Delta
\int_{|k_i|\leq {\frac{\pi}{2}}}{\frac {d {\bf k}}{\pi^2}}
{\frac{1}{E({\bf k})}}
\label{eq:self}
\end{eqnarray}
{}From this expression we arrive to the {\em gap equation}
\begin{equation}
\Delta =2J \lambda \Delta
\int_{|k_i|\leq {\frac{\pi}{2}}}{\frac {d {\bf k}}{\pi^2}}
{\frac{1}{E({\bf k})}}
\label{eq:gap}
\end{equation}

We now discuss the properties of the spectrum found in the AFA.
For general values of $\lambda$, the spectrum of Eq.~(\ref{eq:E_k}) has
a gap $E_g = 4 J\lambda
\Delta $ at the four points in $k $ space $(\pm \pi /2, \pm \pi / 2)$.
Eq.~(\ref{eq:nx}) is a self consistent equation for $\Delta $.
Qualitatively, its
solution as a function of the anisotropy parameter $\lambda $ is a
monotonically increasing function which begins at a critical value of
$\lambda$. From Eq.~(\ref{eq:gap}), it follows that there exists
a critical value of the anisotropy parameter $\lambda_c $ given by
\begin{equation}
{1\over \lambda_{c}}=2
\int_{|k_i|\leq {\frac{\pi}{2}}}{\frac {d {\bf k}}{\pi^2}}
{1\over \sqrt{\cos k_1 ^2 +\cos k_2 ^2 }}
\simeq 2 \times 1.285= 2.570 \ \ ,
\end{equation}
below which $\Delta =0$,
and the spectrum is gapless. Thus, the AFA predicts that the spectrum of
fermions has a gap above a critical anisotropy $\lambda_c \simeq 0.39$.
For small and
positive values of $\lambda-\lambda_c $, the gap has the dependence
\begin{equation}
\Delta \simeq {\rm const.} (\lambda-\lambda_c)
\label{eq:crit}
\end{equation}
and it vanishes for {\em all} $\lambda \leq \lambda_c$.

The vanishing of the gap for $\lambda \to \lambda_c$, and in
particular the exponent of eq.~(\ref{eq:crit}), results from the
collapse of the Fermi surface to four Fermi points and from the linear
dispersion near the Fermi points. Since the density of one-particle
states vanishes in the middle of the band ($E=0$) for a system with a
linear,
relativistic-like, energy-momentum dispersion in two space dimensions,
all instabilities are pushed to finite values of the coupling constants
and there is a critical coupling.
In contrast, in conventional mean field theories of interacting fermions
at finite density on a lattice (but with zero gauge flux) there is no
critical
coupling and the spectrum is always gapped in the presence of nesting.
In such cases, the gap dependence for
small $\lambda $ would be of the form $ \Delta \sim e^{-{1\over
\lambda}}$.
Thus, the existence of the critical value $\lambda_c $ is a
consequence
of the inclusion of the gauge flux that removes the van Hove singularity
in the density of states characteristic of the two dimensional square
lattice.

The exponent of eq.~(\ref{eq:crit}) is valid only at the level of the
AFA. In critical phenomena, it is usually the case that the exponents
found at the level of ``classical" approximations  such as mean field
theory or in the large-$N$ limit, get modified due to the effects of
fluctuations. The AFA is a semiclassical approximation.
Since the dimensionality of space-time of this system is
$2+1$ we should expect non-classical behavior. In fact, the exponent of
eq.~(\ref{eq:crit}) would be correct for a theory of $N$ species of
self-interacting relativistic fermions in the $N \to \infty$ limit. We
will see in section \ref{sec-critical} that the system that we are
studying here is indeed related to a theory of self interacting
relativistic fermions but {\em not} in the large $N$ limit. Thus, in principle,
fluctuations are expected to correct this exponent. However, we will also show
that this second order phase transition is never reached and that it is
preempted by a
fluctuation induced first order transition at a value of lambda
$\lambda^*$ strictly
larger than $\lambda_c$. We will also show that, at this fluctuation induced
first order transition,the system has the expected physical properties of the
isotropic Heisenberg antiferromagnet in a Ne{\`e}l state.

When applied to the isotropic case
($\lambda=1$), the results of the AFA imply a Ne{\`e}l state with a value for
the order parameter $\Delta=0.442 $ and the energy $E=-0.314
J$ per bond. These results should be compared with the
best numerical estimate of the energy, $-0.334J$ per
bond\cite{vmc,huse}. Nevertheless, it is interesting that this
approximation yields a Ne{\`e}l state instead of a flux phase, as one
might have guessed beforehand.

But, is this the correct spectrum? According to the AFA,
the spectrum consists of free fermions with an energy gap for
$\lambda>\lambda_c$
but gapless otherwise. Clearly this spectrum has nothing to
do with
what we found semiclassically in section \ref{sec-1/s}. Earlier work in
anyons and in the FQHE suggests that fluctuations should play a
crucial role. However, notice that since for $\lambda<\lambda_c$
the AFA spectrum is gapless, the fluctuations may yield much more
important effects than what we have described with the AFA. In
particular, this gapless case was found in reference\cite{anyons} in the
case of semions at half filling and was found that the state was not
determined by the AFA alone.

Hence, the next logical step is to look at the effects of fluctuations
around the solutions of the AFA. While it is possible to carry out this
calculation and to keep the full lattice effects at the same time, the
expressions are very cumbersome and not amenable to an analytic
treatment. Instead, we will resort to a different approach in which only the
low energy degrees of freedom are kept. This is equivalent to replace
the system by an effective continuum field theory. These methods are
accurate provided that the gaps do not become too large, in which case
the lattice effects may become dominant. This limitation will force us
to work at values of $\lambda$ close to $\lambda_c$. Nevertheless,
our results will be qualitatively correct even away from this regime.
This approach is pursued in section\ref{sec-critical}.

\section{ Effective Field Theory near $\lambda_c$}
\label{sec-critical}

In this section we will consider the behavior of the system near the
critical anisotropy $\lambda_c$. The mean field theory of section
\ref{sec-meanfield}
yielded an energy gap for the {\it fermions} which vanishes
linearly as $\lambda \to \lambda_c$ from above.
The problem that we
want to address here is the nature of this
phase transition. We
will find that the transition
at $\lambda_{\rm c}$ does not take place as a result of radiative
corrections, namely of fluctuations of the gauge field. Instead, a
different transition with a {\em larger} critical value $\lambda^*$
will
actually occur and supersede the AFA-predicted transition at
$\lambda_{\rm c}$. By counting Goldstone modes we will be able
to identify the transition at $\lambda^*$ with the isotropic
Heisenberg point.

We will now develop an effective theory for the low
energy
degrees of freedom and construct an effective field theory for the
fluctuations about the AFA. To do so we will adopt a point of
view which follows quite closely the treatment of the one-dimensional
anisotropic spin-$1/2$ Heisenberg antiferromagnetic chain\cite{luther}.
However, the physics that we find is very different.
The Jordan-Wigner approach that we have used in this paper
closely resembles the Jordan-Wigner transformation of one-dimensional
systems. The main difference is that, in one space dimension, the
only flux that can be defined is the one that is trapped by the {\it
entire} chain and, hence, it is equivalent to a boundary condition. In
two
dimensions, in addition to global or topological flux, it is possible to
generate local flux. Local fluxes cannot be reduced just to  boundary
conditions and a local, fluctuating, gauge degree of freedom appears
necessarily in the effective theory.

We begin by first summarizing the
standard procedure used in one-dimensional systems, first introduced by
Luther\cite{luther}. We will next follow that construction for the two
dimensional case and use it to discuss the critical behavior near
$\lambda_c$.

The fermionization of the one-dimensional spin chains is usually done in
the following manner\cite{liebmattis}. First, the algebra of spin
one-half operators is realized in terms of canonical fermion operators
{\it via} the use of the Jordan-Wigner transformation. The resulting
system
consists of a set of {\it spinless} fermions hopping between the nearest
neighboring sites of a one dimensional chain of atoms. Two fermions
cannot share the same site (Pauli principle) and only interact when on
nearest neighboring sites with a coupling constant equal to twice the
strength of the $S_z S_z$ coupling constant. The total fermion
number is equal to the number of down spins (depending on the
definitions). There are some subtleties concerning boundary
conditions which are important but are not related to the issues
that are being discussed here. Thus, the sector of the spin system with
total $S_z=0$
maps onto the half-filled portion of the Hilbert space of the Fermi
system. Notice that in one dimension the fermion degrees of freedom
exhaust the Hilbert space and no other degrees of freedom are needed to
represent the states of the spin chain.
 The second step\cite{luther} consists of finding a quantum field
theory in one space dimension which yields the exact long distance
critical behavior of this system of interacting spinless
fermions\cite{emery}.
This is done by separating the slow from the fast components of the
Fermi fields. In one dimension, the fermions can either move to the left
(left movers) or to the right (right movers). The non-interacting Fermi
system (equivalent to the spin one half chain) has two Fermi points with
momenta $p=\pm p_F=\pm \frac{\pi}{2}$. The left and right moving
components of the Fermi field can be thought of as the {\it chiral}
components of a two-component Dirac spinor in one space and one time
moving at a ``speed of light" equal to the Fermi velocity $v_F$. The
equivalent
quantum field theory has the left and right movers interacting through a
backscattering process (up to umklapp processes which are crucial to
reproduce the correct behavior of the isotropic
Heisenberg model\cite{duncan}). The resulting field theory
is the well known Luttinger-Tomonaga-Thirring model\cite{lieb}
which, by using bosonization, can be shown to be equivalent to the
sine-Gordon field theory\cite{luther} (if the umklapp terms are kept).
The result is that for weak Ising coupling ({\it i.e.}\ strong $XY$
anisotropy) the spectrum of the system is that of fermions with
anomalous dimensions (as a result of the backscattering interactions) up
to a value of the backscattering coupling constant at which the umklapp
processes become marginal operators (and beyond which, in the Ising
phase, they are relevant). In the Ising phase there is an energy gap for
all excitations.

In the case of the two-dimensional spin system, the AFA of section
\ref{sec-meanfield}
yielded a spectrum of free fermions with a ``flux phase" band structure.
At half filling ,{\it i.e.}\ total $S_z=0$, the flux phase band structure has
a ``Fermi surface" which reduces to four Fermi points located at
$(\frac{\pi}{2} , \frac{\pi}{2})$ , $(-\frac{\pi}{2} , \frac{\pi}{2})$ ,
$(\frac{\pi}{2} ,-\frac{\pi}{2})$ and $(-\frac{\pi}{2}
,-\frac{\pi}{2})$. Just as in the case of the one-dimensional spin
chain, whose fermion description has two Fermi points, the
two-dimensional problem can also be mapped onto a Dirac-like
problem of Affleck and Marston \cite{affleck}.
The main
differences between the problem that we discuss here and the flux phase
of Affleck and Marston
is that (a) the fermions here are spinless and (b) instead of a
Hubbard-Stratonovich field which fluctuates both in amplitude and phase
we have just a fluctuating phase on the bonds. The
Chern-Simons term is not present in the system discussed
by Affleck and Marston.

We now follow the methods and notations of reference \cite{fradkin2} to
obtain an effective continuum theory. Since the detailed derivation is
rather tedious we will only highlight the procedure. Our general
strategy will be to look for the terms in the action which involve
only low energy degrees of freedom. We will only keep terms with the
smallest numbers of derivatives in each of the fluctuations since
higher derivative terms are irrelevant in the renormalization group
sense. We should keep in mind that this procedure will only give an
approximate value for the coupling constants of the effective low
energy theory since instead of integrating out the high energy degrees
of freedom we are simply neglecting them. Hence, even though the form of
the effective action will still be correct the values of the parameters
({\it e.g.}\ the anisotropy $\lambda$) at which transitions may occur
will not
coincide with the predictions of the lattice system. In particular the
critical anisotropy will not be precisely $\lambda_{\rm c}=1$, but only
approximately. We will have to use the symmetry properties of the
spectrum of excitations at a certain value of the coupling constant of
the effective theory to identify the isotropic point.

The starting
point is the action of eq.~(\ref{eq:finalaccion}). The flux
phase that we found in Section~\ref{sec-meanfield}
has a spectrum of fermions which become gapless at four Fermi points.
The physically important states are those close to the Fermi points. Out
of these states, we construct two (two-component) Dirac
spinors. We begin by defining first a set of spinor components on the
sites
of the real square lattice. It is convenient to split the square lattice
into four sublattices 1, 2, 3 and 4. Sublattice 1 is the set of sites
of the form $\{ {\bf x}=(2n_1,2n_2) \}$ (with $n_1$ and $n_2$ arbitrary
integers). Sublattices 2, 3 and 4 are the sites of the form
$\{ {\bf x}+{\hat {\bf e}}_1 \}$, $\{ {\bf x}+{\hat {\bf e}}_2 \}$ and
$\{ {\bf x}+{\hat {\bf e}}_1 +{\hat {\bf e}}_2\}$ respectively. The
fermion amplitudes on each sublattice will be denoted by $\psi_a({\bf
x})$, with $a=1,\ldots,4$. Likewise, the gauge fields have to be split
into components. This is so because the gauge fields can , and do,
couple
the different fermionic components. In this fashion we will be left with
only slowly varying fields. Thus, with the same notation used for the
fermions, out of the components of the gauge
field ${\cal A}_\mu$ we define four sublattice amplitudes ${\cal A}_\mu^a({\bf
x},t)$
($a=1,\ldots,4$). Since the space components ${\cal A}_j$
only enter in the action in exponential hopping amplitudes, it is
convenient to define
the sublattice amplitudes $W_j^a({\bf x},t)\equiv \exp (i {\cal A}_j^a({\bf
x},t))$. In the uniform flux phase ({\it i.e.}\ for $\lambda < \lambda_{\rm
c}$) the hopping amplitudes take the expectation values
${\bar W}_j^a$ ($j=1,2$). The AFA equations tell us that, in the flux
phase, the oriented
product of the amplitudes ${\bar W}_j^a$ around each plaquette must be
equal to $-1$. In section~\ref{sec-meanfield} we solved this requirement
with the choice of eq.~(\ref{eq:gauge}). For the purposes of taking the
continuum limit, we will choose instead
${\bar W}_1^a=i$ and ${\bar W}_2^a=i (-1)^a$. The two configurations are
related by a gauge transformation and are equivalent modulo a flux of $2
\pi$. The time components have
zero average. It is natural to define a set of slow, fluctuating, fields
${\cal A}_\mu$ by identifying $W_j^a({\bf x},t) \to {\bar W}_j^a \exp (i
{\cal A}_j^a({\bf x},t))$ (notice that the ${\cal A}_j^a$ is now a
fluctuation!).

These fluctuating fields are small but not slow, that is the
fluctuations involve low-frequency processes with wavevectors which are
not necessarily small. In fact, the gauge
fields have components with wave vectors which mix all the fermionic
components. For this reason, we proceed to define a set of fields which
are slow and for which there is a sensible long distance limit. This we
do for both fermions and gauge fields. The sublattice fermion
amplitudes can be combined into two species of Dirac spinors
$\Psi_\alpha^r$
(labelled by $r=1,2$ , each with two components labelled by
$\alpha=1,2$), defined by the following linear combinations
\begin{eqnarray}
\Psi_1^1&=&{\frac{1}{2a_0}} (\psi_1+\psi_2-i \psi_3-i \psi_4)
\nonumber\\
\Psi_2^1&=&{\frac{1}{2a_0}} (-i\psi_1-i\psi_2+\psi_3+\psi_4)
\nonumber\\
\Psi_1^2&=&{\frac{1}{2a_0}} (-i\psi_1+i\psi_2+\psi_3-\psi_4)
\nonumber\\
\Psi_2^2&=&{\frac{1}{2a_0}} (\psi_1-\psi_2-i \psi_3+i \psi_4)
\label{eq:Psi}
\end{eqnarray}
where $a_0$ is a lattice spacing. The normalization factor is
chosen so that the term in the action which includes the time
derivative has coefficient one in the continuum limit. The Dirac
structure is a consequence of the spectrum of the flux phase which is
linear. Unlike the
amplitudes $\psi_a$ which are dimensionless, the continuum fermi fields
$\Psi_\alpha^r$ have dimensions of $(\rm length)^{-1}$. This is the
correct canonical dimension for a Dirac field in $2+1$ dimensions.
We also need to define a set of gamma matrices which in $2+1$ dimensions
are $2 \times 2$ matrices which act on the Dirac components of the
spinors, labelled by the greek index $\alpha$.
We choose them to
be $\gamma_0=\sigma_2$, $\gamma_1=i\sigma_1$ and $\gamma_2=i\sigma_3$,
where $\sigma_i$ ($i=1,2,3$) are the $2 \times 2$ Pauli matrices. We
will also need a second set of Pauli matrices, that we denote by
$T^b$ ($b=1,2,3$), which will act on the species index of the spinors,
labelled by the latin index $r$. When necessary, we will also use the
identity matrix $I$ for each set of indices. In order to simplify the
notation, we will avoid the explicit use of the indices and, instead,
use the matrices to denote the operators of interest.

With these definitions, and after rescaling the time coordinate $t$ by
the Fermi velocity $v_{\rm F}$ as $t=x_0/v_{\rm F}$
(in units of
$J$, we get $v_{\rm F}= a_0$;
here $v_{\rm F}$ is measured in units in which the $XY$ term of
the Hamiltonian has amplitude ${\frac {1}{2}}$), the free part
of the fermion action becomes, in the continuum limit
\begin{equation}
S_{\rm F}^{(0)}=\int d^3x \; {\bar \Psi}_r \; i \slp \; \Psi_r
\label{eq:free} \end{equation}
with ${\bar \Psi}=\Psi^* \gamma_0$. This continuum theory is valid for
fluctuations with wavevectors smaller than some cutoff $\Lambda \approx
{\frac {\pi}{a_0}}$, where $a_0$ is the lattice spacing. The fact that
this is an effective theory at length scales long compared with the
lattice spacing will have important consequences for our analysis. In
field theory the choice of cutoff (or regularization) is largely
arbitrary. Different choices of regularization usually lead to the same
theory. However, for theories of fermions a number of subtleties arise
connected with the way regularizations treat the symmetries of the
effective continuum theories. In $2+1$ dimensions relativistic massive
fermions have a parity anomaly. However, if the fermion mass has a
dynamical origin ({\it i.e.}\ if it is induced by fluctuations) the parity
anomaly may be lost in some regularization methods (such as
dimensional regularization) which set to zero all non logarithmic
divergencies.
In any case, in our problem we are not free to choose an arbitrary
regularization
scheme to cutoff the divergencies present in various Feynman diagrams
of the effective continuum theory.
Instead the choice of cutoff will be done in such a way
that the symmetries of the {\em lattice} system are respected.
In particular, we will see in the next section that schemes such as
dimensional regularization cannot be used for this problem.

Next we define new linear combinations of the gauge fields. We choose
the new linear combinations which have a dominant wavevector $(q_1,q_2)$
such that it mixes the fermionic components defined above. We use the
notation $A_\mu^{q_1 q_2}$. The important wavevectors are
${(q_1,q_2)}=(0,0),(\pi,0),(\pi,\pi),(0,\pi)$. The new fields are
\begin{eqnarray}
A_\mu &=&
{\frac {1}{4a_0}}({\cal A}_\mu^1+{\cal A}_\mu^2+{\cal A}_\mu^3+{\cal A}_\mu^4)
\label{eq:A_mu}\\
A_\mu^{\pi 0}&=&
{\frac{1}{4a_0}}({\cal A}_\mu^1-{\cal A}_\mu^2+{\cal A}_\mu^3-{\cal A}_\mu^4)
\label{eq:A_mu^{pi 0}}\\
A_\mu^{\pi \pi}&=&
{\frac {1}{4a_0}}({\cal A}_\mu^1-{\cal A}_\mu^2-{\cal A}_\mu^3+{\cal A}_\mu^4)
\label{eq:A_mu^{pi pi}}\\
A_\mu^{0 \pi}&=&
{\frac {1}{4a_0}}({\cal A}_\mu^1+{\cal A}_\mu^2-{\cal A}_\mu^3-{\cal A}_\mu^4)
\label{eq:A_mu^{0 pi}}
\end{eqnarray}
where $A_\mu \equiv A_\mu^{00}$ are the smooth fluctuations of the
gauge field.

The only fluctuations that are usually kept in
these type of analysis are the smooth fields like $A_\mu$. However, we
find that some of
the other amplitudes are very important. In fact we will find that the
fluctuations $A_\mu^{\pi \pi}$ are connected with Ne{\`e}l
fluctuations.
The fields $A_\mu^{q_1 q_2}$
couple to operators which are bilinear in Fermi fields and whose
characteristic wavevector is $(q_1,q_2)$.

In terms of the
fermion amplitudes defined in eq.~(\ref{eq:Psi}) there
is a total of $16$ operators which are bilinear
combinations of the amplitudes $\Psi_\alpha^r$. They have
the form $M={\bar \Psi} \Psi$, $M^b = {\bar \Psi} T^b \Psi$,  $M_\mu
={\bar \Psi} \gamma_\mu \Psi$ and $M_\mu^b ={\bar \Psi} \gamma_\mu T^b
\Psi$.
In spin language these fermion bilinears correspond to
linear combinations of site occupation numbers and of hopping
amplitudes among sites on the four sublattices. Out of the $16$
bilinears we will only discuss
three of them, $M$, $M^3$ and $M_0$, which correspond to the parity
breaking mass
operator, the Ne{\`e}l order parameter and the charge density operator
respectively. We find the following identifications
\begin{eqnarray}
M={\bar \Psi} \Psi & \leftrightarrow &
i(\psi_1^\dagger \psi_4-\psi_4^\dagger \psi_1)+
i(\psi_3^\dagger \psi_2-\psi_2^\dagger \psi_3)
\label{eq:chiral}\\
M_0={\bar \Psi} \gamma_0 \Psi & \leftrightarrow &
\psi_1^\dagger \psi_1+\psi_2^\dagger \psi_2+
\psi_3^\dagger \psi_3+\psi_4^\dagger \psi_4
\label{eq:charge}\\
M^3={\bar \Psi} T^3 \Psi & \leftrightarrow &
-\psi_1^\dagger \psi_1+\psi_2^\dagger \psi_2+
\psi_3^\dagger \psi_3-\psi_4^\dagger \psi_4
\label{eq:Neel}
\end{eqnarray}
These identifications show that $M_0$ is the total fermion occupation
number averaged over the four sublattices. In the sector with $S_z=0$ we
expect to find $\langle M_0 \rangle =0$.  The operator $M^3$ is the
staggered occupation fermion number which, back in spin language is the
Ne{\`e}l order parameter. Finally, $M$ is an operator which induces
hopping across the main diagonals of the plaquettes. It is also easy to
show that the phase factors
present in the definition of $M$, when combined with the phase factors
of the flux phase, indicate that the flux on {\em every} elementary {\em
triangle} inscribed in each plaquette is equal to $\frac{\pi}{2}$. Thus,
a non-zero value of $\langle M \rangle$ in the ground state breaks both
$T$ and $P$. $M$ is the Chiral Order parameter introduced by Wen,
Wilczek and Zee\cite{wwz}. We will see in section\ref{sec-phase} that
this operator plays a very important role in the determination of the
physically correct ({\sl PC\/}) phase diagram.

The next step is to write the
action of eq.~(\ref{eq:finalaccion})
in terms of the slowly varying fields $\Psi_\alpha^r$ and
$A_\mu^{q_1q_2}$. We discuss first the Fermion part of the
effective Lagrangian ${\cal L}_{\rm F}$ in the continuum limit. After
expanding the hopping amplitudes up
to leading order in fluctuations and after taking the continuum limit we
get
\begin{equation}
{\cal L}_{\rm F}=
{\bar \Psi} \; i \slD \; \Psi +
A_0^{\pi \pi} {\bar \Psi} T^3 \Psi
\label{eq:L_F}
\end{equation}
where $\slD \equiv D_\mu \gamma^\mu$ and $D_\mu= \partial_\mu -i A_\mu$
is the covariant derivative which represents the coupling to the long
wavelength smooth pieces of the Chern-Simons gauge field. We have not
included in the final form of the Lagrangian, additional terms
of the form ${\cal L}_{\rm extra}$,
\begin{equation}
{\cal L}_{\rm extra}=
A_0^{0 \pi} {\bar \Psi} T^2 \gamma_1\Psi +
A_0^{\pi 0} {\bar \Psi} T^1 \gamma_2 \Psi -
A_1^{0 \pi} {\bar \Psi} T^2 \gamma_0 \Psi -
A_2^{\pi 0} {\bar \Psi} T^1 \gamma_0 \Psi
\end{equation}
which couple the fermions to fluctuations with wavevectors $(0, \pi)$
and $(\pi,0)$, since it is possible to show that they are irrelevant in
the renormalization group sense. Physically, this can be understood
since these operators do not acquire an expectation value in any of the
phases of our interest. Using the linear combinations
of eq.~(\ref{eq:Psi}), one can show that these operators are
related with
both spin density wave order parameters with wave vectors $(0 \pi)$ and
$(\pi 0)$ and to Peierls (or dimer) order parameters with the same wave
vectors. Neither type of order occurs in this system.  From now on we
will ignore these terms. Notice that these terms have the same scaling
dimension as the ones that are being kept. It is conceivable that there
are other situations in which these terms become dominant such as in the
vicinity of a dimerization transition. Such phase transitions are
possible in a frustrated antiferromagnet. However, the Chern-Simons
( or Wigner-Jordan) mapping used here cannot be used as it stands for
non-bipartite lattices. The terms in ${\cal L}_{\rm extra}$
break the rotational invariance of the continuum theory of
eq.~(\ref{eq:L_F}) down to the allowed symmetries of the point group
of the square lattice. Again, such anisotropies will only be relevant in
the vicinity of ground states which break such symmetries. Below we will
also ignore terms in the bosonic part of the Lagrangian with the same
symmetry properties.

Finally, we need to find the continuum form of the bosonic parts of the
action (\ref{eq:finalaccion}). We only keep terms to leading order in
the lattice spacing for each field. There are two sets of contributions.
One set comes from $S_{\rm int}$. When written in terms of the slow
components of the gauge fields they contribute with the term ${\cal L}_{\rm
int}$
\begin{equation}
{\cal L}_{\rm int}={ \frac{g}{2}} \left(A_2^{\pi \pi} - A_1^{\pi \pi}
\right)^2
-{\frac {1}{2 {\bar e}^2}} \left(\partial_1 A_2- \partial_2 A_1\right)^2
\label{eq:Lint}
\end{equation}
where the effective coupling constants $g$ and ${\bar e}^2$ are
\begin{equation}
\begin{array}{ccc}
g={\frac {4 \lambda}{\pi^2 a_0}} & {\;\;\;\;\;} & {\bar e}^2=
{\frac{\pi^2}{\lambda a_0}}
\label{eq:coupling}
\end{array}
\end{equation}
The Maxwell-like term in the Lagrangian
${\cal L}_{\rm int}$ has no consequence on the phase diagram since it
has one
more derivative than the Chern-Simons term and, hence, it is irrelevant
at long distances~\cite{maxwell} and it will be ignored from now on.

The second set of contributions comes from the Chern-Simons terms
$S_{\rm cs}$. We find a contribution to the Lagrangian of the form
\begin{equation}
{\cal L}_{\rm cs}=
{\frac {\theta}{4}} \epsilon^{\mu \nu \lambda} A_\mu
F_{\nu \lambda} +
{\frac {2 \theta}{a_0}}A_0^{\pi \pi}\left(A_1^{\pi \pi}- A_2^{\pi
\pi}\right)
\label{eq:Lcs}
\end{equation}
where $F_{\mu \nu}=\partial_\mu A_\nu-\partial_\nu A_\mu$ is the field
strength.

Before putting everything together we notice, by inspecting
eqs.~(\ref{eq:Lint}) and (\ref{eq:Lcs}), that the fields
$A_1^{\pi \pi}$ and $A_2^{\pi \pi}$ enter in the total Lagrangian in
terms of the difference $A_1^{\pi \pi}-A_2^{\pi \pi}$ and that the
total Lagrangian is at most quadratic in this difference. Thus, we can
integrate out these fields and find the Lagrangian of
the form
\begin{equation}
{\cal L}=
{\bar \Psi} \; i \slD \; \Psi +
A_0^{\pi \pi} {\bar \Psi} T^3 \Psi - {\frac{1}{2 {\bar g}}}
\left(A_0^{\pi \pi} \right)^2+
{\frac {\theta}{4}} \epsilon^{\mu \nu \lambda} A_\mu F_{\nu \lambda}
\label{eq:L}
\end{equation}
where ${\bar g}$ is an effective coupling constant. It is convenient to
define a dimensionless coupling constant $g_0$ such that ${\bar g}=g_0
a_0$. In terms of the anisotropy $\lambda$ and of the Chern-Simons
coupling $\theta$ we get
\begin{equation}
g_0= {\frac{\lambda}{\pi^2 \theta^2}}=4 \lambda
\left(\frac{\delta}{\pi}\right)^2
\label{eq:g_0}
\end{equation}
where we have introduced the statistical angle $\delta=1/2\theta$.
Notice that for $\theta\to {\frac{1}{2\pi}}$ ({\it i.e.}\ in the boson limit of
interest here) $\delta \to \pi$ and $g_0 \to 4 \lambda$. Conversely, in
the fermion limit $\delta \to 0$ or $\theta \to \infty$, the
dimensionless coupling constant is weak, $g_0 \to 0$.
Thus, in the hard-core boson limit, which is the case of interest, we
are dealing with a system in which the dimensionless coupling
constant is typically of order unity and perturbation theory should not
be reliable.

Let us discuss the physical meaning of the terms of the Lagrangian ${\cal L}$
of eq.~(\ref{eq:L}). The field $A_0^{\pi \pi}$ couples locally to the
Ne{\`e}l order parameter operator $M^3={\bar \Psi} T^3\Psi $.
Thus, if
$A_0^{\pi \pi}$ picks up an expectation value, so does the order
parameter $M^3$. Clearly such a state has Ne{\`e}l long range order. In
section \ref{sec-meanfield} we found that beyond some critical value of
the anisotropy $\lambda$ the system is in a Ne{\`e}l state. The field
$A_0^{\pi \pi}$ also enters at most quadratically in the Lagrangian
${\cal L}$ and it can also be integrated out (in fact, we may regard the
field $A_0^{\pi \pi}$ as a Hubbard-Stratonovich field). Indeed, after
integrating out $A_0^{\pi \pi}$ the total Lagrangian ${\cal L}_{2{\rm D}}$ for
the two-dimensional system has the suggestive form
\begin{equation}
{\cal L}_{2{\rm D}}=
{\bar \Psi} \; i \slD \; \Psi + {\frac{\bar g}{2}}
\left({\bar \Psi} T^3\Psi \right)^2        +
{\frac {\theta}{4}} \epsilon^{\mu \nu \lambda} A_\mu
F_{\nu \lambda}
\label{eq:L_2D}
\end{equation}
Thus, we find that the low energy degrees of freedom of the anisotropic
quantum antiferromagnet can be described in terms of a theory of two
relativistic Fermi fields in $2+1$ dimensions, with a
four-Fermi interaction of strength ${\bar g}$, and which are also
coupled with a Chern-Simons gauge
field. The Chern-Simons coupling constant is restricted to the value
$\theta={\frac{1}{2 \pi}}$.
For the rest of this paper, we will consider this Lagrangian.

In the form of eq.~(\ref{eq:L_2D}), the effective continuum theory for
the two dimensional
system is a generalization of the equivalence that exists between the
one-dimensional anisotropic spin-${\frac {1}{2}}$ quantum Heisenberg
antiferromagnet and a theory of a self-interacting relativistic Fermi
field, the Luttinger-Thirring
model\cite{luther,duncan}. The left and right
components of the Fermi fields, $\Psi_{\rm L}$ and $\Psi_{\rm R}$, can
be
viewed as the two components of a Dirac spinor in $1+1$ dimensions. The
Lagrangian ${\cal L}_{1 {\rm D}}$ of the effective field theory for the
antiferromagnetic spin chain is \cite{gn}
\begin{equation}
{\cal L}_{1 {\rm D}}=
{\bar \Psi} \; i \slp \; \Psi + {\frac{{\bar g}_{1D}}{2}}
\left({\bar \Psi} \Psi \right)^2
\label{eq:L_1D}
\end{equation}
The coupling constant for the 1D system is ${\bar g}_{1D}=2 \lambda$
which is dimensionless.

The two-dimensional theory, with Lagrangian ${\cal L}_{2{\rm
D}}$, and the one-dimensional theory with Lagrangian ${\cal L}_{1{\rm D}}$
differ in a number of ways: (a) in
1D there is only {\em one} species of Dirac fermions (instead of the two
labelled by $r$ in ${\cal L}_{2{\rm D}}$ ), (b)
there is {\em no gauge field} in 1D but there is a Chern-Simons gauge
field in 2D, (c) ${\bar g}_{1D}$ is dimensionless while ${\bar g}$
has dimensions of length and (d) the symmetries are different. They both
have a self-interacting, quartic, term
in fermions which, in both cases, is equal to the square of the Ne{\`e}l
order parameter.
We will also see that the coupling to the Chern-Simons
term is, in this problem, responsible for much more than a change in
statistics.

\section{Critical Behavior and Phase Diagram}
\label{sec-phase}

In this section we discuss the phase diagram of the effective field
theory derived in section\ref{sec-critical} with the Lagrangian
${\cal L}_{2{\rm D}}$ of Eq.~(\ref{eq:L_2D}).
Given the apparent similarity
between this theory and its analog in one dimension  ${\cal L}_{1{\rm D}}$
of Eq.~(\ref{eq:L_1D}), one might think that the phase diagrams may also
be quite similar. However, a closer analysis shows that this cannot
possibly be correct.

The two theories have different symmetries as well as different scaling
properties. In both cases, the order parameter of the Ne{\`e}l state is
odd under a
sublattice exchange or, what is the same, it is odd under a global shift
of the field configuration by one lattice spacing. This is a global
discrete symmetry which can be spontaneously broken by the ground state,
even in one space dimension. In the case of the one dimensional chain,
the Ne{\`e}l order parameter, {\it i.e.}\ the difference of the spinless
fermion occupancy between the two sublattices, is proportional to
${\bar \Psi} \Psi$.
This operator is odd under the transformation $\Psi \to \gamma_5 \Psi$
which changes the relative sign of the right and left moving amplitudes
of the fermions. This symmetry is known as a discrete chiral symmetry.
The Lagrangian
${\cal L}_{1 {\rm D}}$ is even under this chiral symmetry. As a result,
it is possible to show that the operator ${\bar \Psi} \Psi$ does not
acquire an expectation value to all orders in perturbation theory in the
coupling constant ${\bar g}_{1D}$. Also, because this symmetry is
present, renormalization effects do not induce fermion mass terms
in the Lagrangian, which are proportional to ${\bar \Psi} \Psi$ and
thus brake the symmetry explicitly.

The $1+1$-dimensional system has very special scaling properties. In
space-time dimensions $D=2$, the coupling constant ${\bar g}_{1D}$
is dimensionless. Then, the standard field-theoretic analysis tells us
that the theory defined by the Lagrangian ${\cal L}_{1
{\rm D}}$, the Gross-Neveu model~\cite{gn}, is
renormalizable. If the number of fermionic species is $N$,
this theory is asymptotically free with a $\beta$-function~\cite{sign}
$\beta({\bar g}_{1D})=\left({\frac{N-2}{2 \pi}}\right){\bar g}_{1D}^2$.
For $N
\geq 2$ this $\beta$-function is strictly positive and the system has a
dynamically
generated energy gap.  For the
case of interest for the {\em one-dimensional} quantum Heisenberg
antiferromagnet, $N=1$ and
the coupling constant is still dimensionless. However, in this case, the
$\beta$-function
vanishes to all orders in perturbation theory and the system has a
line of fixed points. In other words, the four fermion operator
$({\bar \Psi} \Psi)^2$ is {\em marginal}. Bosonization studies
show~\cite{luther,dennijs,umpklapp} that there is an
operator which is irrelevant at small coupling but that at a critical
value of the coupling constant (namely for a critical anisotropy) it
becomes relevant. The effect of this operator,
which represents umklapp processes in the lattice theory, is to end the
line of fixed points at a multicritical point which is in the
Kosterlitz-Thouless universality class. An explicit computation of the
critical exponents shows~\cite{luther} that the correlation functions at
this multicritical point have the symmetries of the isotropic Heisenberg
antiferromagnet.

In the case of the
two-dimensional system, the order parameter is ${\bar \Psi} T^3 \Psi$.
Unlike one dimension, there is no $\gamma_5$ Dirac matrix in $D=2+1$.
Instead, the order parameter is now odd under an operator which
effectively exchanges the two species of fermions. However, a mass term
proportional to the operator ${\bar \Psi} \Psi$ is {\em even} under the
exchange of fermionic species (or sublattices, which is the same) but it
is {\em odd} under parity ($P$) and time reversal ($T$) transformations.
In contrast, ${\bar \Psi} T^3 \Psi$ does not break these symmetries. The
reason is that, for {\em each} species, the operator ${\bar \Psi} \Psi$
changes sign under $P$ and $T$ and, hence, ${\bar \Psi} T^3 \Psi$
changes sign too. However, this effect is equivalent to a redefinition
of the sublattices and, thus, it breaks neither parity nor time
reversal. For this reason, the operator ${\bar \Psi} T^3 \Psi$ is
usually called a parity invariant mass term. Thus, renormalization
effects cannot induce a parity invariant mass term in perturbation
theory since it is a symmetry breaking field. We will show below that
the phase transition found in section \ref{sec-meanfield} represents the
spontaneous breaking of the symmetry ${\bar \Psi} T^3 \Psi \to -{\bar
\Psi} T^3 \Psi $ to a Ne{\`e}l phase in which $\langle {\bar \Psi} T^3
\Psi\rangle \not= 0$.

Unlike the $1+1$-dimensional theory, the Lagrangian
${\cal L}_{2 {\rm D}}$
of eq.~(\ref{eq:L_2D}) has, in addition to the discrete
symmetry ${\bar \Psi} T^3 \Psi \to {\bar \Psi} T^3 \Psi $, a
continuous gauge (local) symmetry,
\begin{eqnarray}
\Psi(x) & \to & e^{i \phi(x)} \Psi (x) \nonumber \\
A_\mu(x) & \to & A_\mu(x)+ \partial_\mu \phi(x)
\label{eq:gaugesymmetry}
\end{eqnarray}
With the rules that we are using here, the gauge field has
scaling dimension one and the Chern-Simons term has scaling dimension
three and it is marginal. The Chern-Simons coupling constant $\theta$ is
dimensionless and the theory is renormalizable. In contrast, the four
fermion operator has na{\"\i}ve scaling  dimension
four and it is irrelevant. The four fermion coupling constant
${\bar g}$
has na{\"\i}ve scaling dimension of length. The dimensionless
coupling constant $g_0$ of eq.~(\ref{eq:g_0}) has a {\em negative}
$\beta$-function, $\beta(g_0)=-g_0+ O(g_0^2)$ and $g_0^*=0$ is an
infrared stable fixed point of the renormalization group flow. In
contrast, the Chern-Simons coupling $\theta$ has a vanishing
$\beta$-function. Thus, at least na{\"\i}vely, it appears that this
renormalization group flow has a line of fixed points parametrized by
$\theta$.

W.~Chen and M.~Li\cite{weichen} have recently put
forward arguments in favor of a scenario in which theories of
relativistic
fermions coupled to Chern-Simons gauge fields have a line of fixed
points parametrized by $\theta$. In particular they argue
that the fluctuations of the gauge fields induce anomalous dimensions in
the fermion operators and, as a result, the four-fermi operators become
less irrelevant as $\theta$ grows larger. However, their analysis is
based in dimensional regularization. It is well known that this
regularization scheme sets to zero all singular Feynman diagrams
except for logarithmically divergent terms. Since it is an analytic
regularization method, it also sets to zero all contributions which
cannot be continued in dimension. Thus, fermion mass terms are not
induced in dimensional regularization.

At the infrared stable fixed point ${\bar g}=0$, the Lagrangian ${\cal
L}_{2 {\rm D}}$ is manifestly scale invariant. The only parameter left
is the Chern-Simons coupling constant $\theta$ which is dimensionless.
Hence, at least in the absence of fluctuations, it is at a fixed point
and it has no scale. However, it is easy to convince oneself that
fluctuations lead to divergent corrections of the classical (that is,
mean field) values of the observables. The presence of divergent
contributions
in the perturbation series requires the use of a cutoff or, more
generally, of a regulator. Any cutoff introduces a microscopic scale in
the problem and therefore it breaks the apparent scale invariance. Thus,
in the presence of a cutoff, dimensionful terms can be induced by
renormalization effects. On dimensional grounds,
an induced mass term will have to be proportional to the cutoff since
there is no other scale left.
The theory that we are studying has a natural cutoff, the lattice
spacing, which acts as a natural scale.
Hence, in this case, it is physically incorrect to use dimensional
regularization.
Furthermore, the Chern-Simons term breaks both $T$ and $P$. Thus, it is
expected that renormalization effects should generate all possible terms
which break the same symmetries. The parity-odd fermion mass term ${\bar
\Psi} \Psi$ breaks precisely the same symmetries. Hence, unlike the
$1+1$-dimensional case there is no symmetry that prohibits these terms
to be induced by renormalization.

Unlike the $1+1$-dimensional theory, non-perturbative tools such as
bosonization are not available for the study of relativistic systems in
$2+1$ dimensions. In order to proceed further we will use a
semiclassical approximation, like the one of section
\ref{sec-meanfield}, but going beyond the leading order. In
$1+1$-dimensions this approach would not be sufficient since this
approximation misses the marginality of the interaction. However, in
$2+1$ dimensions the four fermion interaction is irrelevant (in weak
coupling) and this approximation reproduces this result correctly.
In order to determine the induced fermion mass $M_{\rm ind}$, we will
compute the leading self-energy
correction to the fermion propagator due to fluctuations of the gauge
field. Since the
gauge field couples in the same way to both species of fermions, the
only
possible induced mass term is ${\bar \Psi} \Psi$ which is even under the
exchange of species. Since this term is odd under $T$ and $P$ it can
only arise from fluctuations of the Chern-Simons gauge field.

The partition function at zero temperature is
\begin{equation}
{\cal Z}=\int {\cal D}{\bar \Psi} \; {\cal D} \Psi \; {\cal D} A_\mu
\; {\cal D} A_0^{\pi \pi} \;
e^{i\int d^3x \; {\cal L}( {\bar \Psi}, \Psi, A_\mu,A_0^{\pi \pi} )}
\end{equation}
where ${\cal L}$ is the Lagrangian of Eq.~(\ref{eq:L}), which we
reproduce here for clarity
\begin{equation}
{\cal L}=
{\bar \Psi} \; i \slD \; \Psi +
A_0^{\pi \pi} {\bar \Psi} T^3 \Psi
- {\frac{1}{2 {\bar g}}} \left(A_0^{\pi \pi} \right)^2+
{\frac {\theta}{4}} \epsilon^{\mu \nu \lambda} A_\mu F_{\nu \lambda}
\label{eq:L'}
\end{equation}
This form of the theory makes the semiclassical approximation more
transparent. We now follow the methods outlined in section
\ref{sec-meanfield} and find an effective action $S_{\rm eff}$ for the
Bose fields,
which here are $A_\mu$ and $A_0^{\pi \pi}$, after integrating out the
fermions:
\begin{equation}
S_{\rm eff}=-i \; {\rm Tr} \ln \left[ i \slD + A_0^{\pi \pi} T^3 \right]
-\int d^3x\; {\frac{1}{2 {\bar g}}} \left(A_0^{\pi \pi} \right)^2+ \int
d^3x \; {\frac {\theta}{4}} \epsilon^{\mu \nu \lambda} A_\mu F_{\nu
\lambda}
\label{eq:effaction}
\end{equation}
The stationary points of this action satisfy the equations
\begin{eqnarray}
{\frac {\delta S_{\rm eff}} {\delta A_\mu (x)}}&=&i \; {\rm Tr}\left[
S(x,x)
\gamma_\mu \right] + {\frac{\theta}{2}} \epsilon_{\mu \nu \lambda}
F^{\nu \lambda}=0 \\
{\frac {\delta S_{\rm eff}} {\delta A_0^{\pi \pi} (x)}}&=&i \; {\rm
Tr}\left[ S(x,x) T^3 \right]- {\frac{1}{ {\bar g}}} A_0^{\pi \pi}(x)=0
\label{eq:contspe}
\end{eqnarray}
where $S(x, x')$ is the (Feynman) fermion propagator
\begin{equation}
S(x,x')=\langle x \vert {\frac{1}{i \slp + A_\mu \gamma^\mu + A_0^{\pi
\pi} T^3}} \vert x' \rangle
\label{eq:contprop}
\end{equation}
In order to simplify the notation we have dropped all the indices that
are attached to the fermions. The traces run over both spinor (Dirac)
and ``flavor" (species) indices.

The equations (\ref{eq:contspe}) have a solution of the form $A_\mu=0$
and $A_0^{\pi \pi}= M$. Please recall that the gauge field $A_\mu$ is
the long wavelength fluctuation of the (lattice) Chern-Simons gauge
field around the flux phase. $M$ is the Ne{\` e}l mass and it is given
by
\begin{equation}
{\frac{M}{{\bar g}}}=i\;\int {\frac {d^3p}{(2 \pi)^3}} {\rm Tr}
{\frac{1}{{\lnp}-M T^3+i \epsilon}} T^3
\label{eq:gapcont}
\end{equation}
which is the continuum analog of the gap equation of section
\ref{sec-meanfield}. After performing the integral (with a cutoff on the
space components of the momentum $\Lambda \approx {\frac {\pi}{2 a_0}}$)
we find that the mass $M$ is the solution of
\begin{equation}
{\frac{ M }{ {\bar g}}}={\frac {\Lambda}{\pi}} \left( {\sqrt{1 + {\frac
{M^2}{ \Lambda^2}}}}-{\frac {\vert M \vert}{\Lambda}} \right) M
\label{eq:newgap}
\end{equation}
This {\em gap equation} has the solution
\begin{equation}
\vert M \vert = {\frac {\Lambda}{2}} \left( {\frac{{\bar g}}{{\bar
g}_c}}-{\frac{{\bar g}_c}{{\bar g}}}\right) \Theta({\bar g}-{\bar g}_c)
\label{eq:M}
\end{equation}
where $\Theta (t)$ is the Heavyside function and ${\bar g}_c=
{\frac{\pi}{ \Lambda}}$
is the critical coupling constant of the effective continuum theory.
Using this value of ${\bar g}_c$ and eq.~(\ref{eq:g_0}) we get an
estimate for the critical anisotropy $\lambda_c$ from this continuum
theory given approximately by $\lambda_c \approx {\frac {1}{2}}$ which
should be compared with the (lattice) AFA value $\lambda_c \approx 0.4$
of section \ref{sec-meanfield}. This different value of $\lambda_c$,
which is
a non-universal quantity, reflects the approximations made in taking the
continuum limit. In particular, it depends on the precise relation
between the momentum cutoff $\Lambda$ and the lattice spacing $a_0$.

Let us
denote the full fermion propagator at momentum $p_\mu$ ($\mu=0,1,2$) by
$S(p)$ and by $G^{\mu \nu}(p)$ the
propagator of the Chern-Simons gauge field.
At this level of approximation,
the fermion propagator, at $3$-momentum $p_\mu$, is given
by
\begin{equation}
S_0(p)= {\frac {{\lnp}+M T^3}{p^2-M^2+i \epsilon}}
\label{eq:fermionprop}
\end{equation}
where, once again, we have dropped all indices.
The bare propagator of the gauge field at $3$-momentum $p_\mu$,
in the Lorentz gauge $\partial_\mu A^\mu=0$, is
\begin{equation}
G^{\mu \nu}_0(p)=
{\frac {i}{\theta}} \epsilon^{\mu \nu \lambda}
{\frac {p_\lambda}{p^2+i \epsilon}}
\label{eq:gaugeprop}
\end{equation}
Thus, in every Feynman diagram, each propagator of the gauge field
contributes with a weight proportional to $1/\theta=2\delta$, where
$\delta$ is the statistical angle. Therefore, this is an expansion in
powers of the statistics and it is accurate only near the fermion limit
$\delta \to 0$ or $\theta \to \infty$. The value of $\delta$ of interest
for the Heisenberg antiferromagnet is $\delta = 2 \pi$ which is not
small. A more serious problem is that the physical properties of systems
of this sort must be periodic in the statistics. Namely, all amplitudes
for any physical observables must not change under the replacements
$\delta \to \delta+ 2 \pi k$, where $k$ is an {\em even} integer
(periodicity) and $\delta \to 4 \pi - \delta$ (symmetry around bosons).
We will use parturbation theory around fermions and demand that it holds
around each {\em period}. The extrapolation to the boson point $\delta=2
\pi$ should yield qualitatively correct results.

The exact fermion propagator $S(p_\mu)$ obeys the Dyson equation
\begin{equation}
S(p)^{-1}=S_0(p)^{-1}-\Sigma(p)
\label{eq:dysonfermion}
\end{equation}
where $\Sigma(p)$ is the fermion self-energy. Due to the symmetry
of the bare theory, $\Sigma(p)$ has the form
\begin{equation}
\Sigma_{rr'}(p)=\Sigma_1(p) +\Sigma_2(p) T^3
\end{equation}
To leading order in $\delta$, $\Sigma(p)$ is given by
\begin{equation}
\Sigma(p)=\int {\frac {d^3k}{(2\pi)^3}} \;
\gamma^\mu S_0(p-k)\; \gamma^\nu i G^{\mu \nu}_0(k) \; +
O(\delta^2)
\end{equation}
Explicitly, we find
\begin{equation}
\Sigma(p)=i \; \int {\frac {d^3k}{(2\pi)^3}} \; \gamma^\mu
{\frac {{\lnp}-{\lnk}+M T^3}{(p-k)^2-M^2+i \epsilon}}
\gamma^\nu
{\frac {i}{\theta}} \epsilon_{\mu \nu \lambda}
{\frac {i \; k^\lambda}{k^2+i \epsilon}}+ O(\delta^2)
\label{eq:ferself}
\end{equation}
By counting powers of the momentum of integration we see that this
contribution has an ultraviolet, linear divergence.
By expanding the self energy in powers of the external momentum $p_\mu$,
we see that the (ultraviolet) linear divergence only affects the term
at zero external momentum $p=0$ and that all contributions at non-zero
external momentum are finite. This ultraviolet divergence is an
artifact of ignoring the fact that we are working with an effective field
theory and that the quantum antiferromagnet, from which this field theory
is derived, is defined on a lattice and
it does  have a cutoff. Thus, this divergence has to be cutoff at values of the
internal momentum of the order of $\Lambda \approx \pi / 2 a_0$ ,
where $a_0$ is the lattice spacing. After some algebra we find
\begin{equation}
\Sigma(p)={\frac{ i}{\theta}} \int_0^1  dx \; \int {\frac {d^3q}{(2
\pi)^3}} {\frac {\left(q_\lambda+x
p_\lambda\right)\left[ \left( p_\rho(1-x)-q_\rho
\right)\; 2 \; g^\rho_\lambda-2M T^3
\gamma_\lambda\right]}{\left[q^2+p^2x(1-x)-M^2x\right]^2}}
\label{eq:integral}
\end{equation}
where we have used the covariant notation $q^2=q_0^2-{\bf q}^2$ and an
$i \epsilon$ prescription is assumed.

After an integration over the
frequency variable $q_0$ and over the spacial components of the momentum
of integration ${\bf q}$ (with a cutoff $\Lambda$) we get
\begin{eqnarray}
\Sigma(p)&=& \int_0^1 dx  \left\{
\left(xMT^3 {\lnp} \right) {\frac{1}{4 \pi \theta}}
 \left[  {\frac{1}{{\sqrt{ M^2x-p^2 x(1-x)}}}}
- {\frac{1}{{\sqrt{\Lambda^2+M^2x-
p^2x(1-x)}}}} \right] \right. \nonumber \\
& &-{\frac{1}{2 \pi \theta}}
 \left[  {\sqrt{\Lambda^2+M^2x- p^2x(1-x)
}}- {\sqrt{M^2x-p^2 x(1-x)}} \right] \nonumber \\
 & & \left. + {\frac{1}{4 \pi \theta}}
\left(M^2x-2x (1-x)p^2\right)
 \left[ {\frac{1}{{\sqrt{M^2x- p^2x(1-x)}}}}
- {\frac{1}{{\sqrt{\Lambda^2+M^2x-p^2
x(1-x)}}}} \right] \right\}
\label{eq:integral2}
\end{eqnarray}

Since we are only interested in the computation of the effective (or
renormalized) mass, it will be sufficient for our purposes to
compute the integral of eq.~(\ref{eq:integral2}) at $p=0$. In this limit
we find that $\Sigma_2(0)=0$ and we get a value for $\Sigma(0)$
which is independent of the fermionic species. In this limit, and after
some algebra, eq.~(\ref{eq:integral2}) becomes
\begin{equation}
\Sigma_{rr'}(0) = -{\frac{\Lambda}{2 \pi \theta}} \left({\sqrt{
1+{\frac{M^2}{\Lambda^2} }}}-{\frac{M}{\Lambda}} \right) \delta_{rr'}
\label{eq:sigma}
\end{equation}
This contribution to the self energy of the fermion
plays the role of an effective or {\em induced}
mass and we will denote it by $\Sigma(0)\equiv M_{\rm ind}$. Due to
the symmetry of the bare fermion propagator, this (divergent) induced
mass is {\em the same}
for all fermionic species which thereby acquire the {\em same} mass.
This result also tells us that the induced mass is proportional to
$-1/\theta$. The fact that this sign is {\em opposite} to the
sign $\theta$ will play a fundamental role in our analysis.

Let us now use these results to  compute the mass of {\em each} species
of fermions up to corrections of order $\delta^2$. Our calculation tells
us that the total fermion propagator $S(p_\mu)$ at zero external
momentum has the form
\begin{equation}
S^{-1}(0)=-M T^3-\Sigma(0)
\end{equation}
from where we find that the effective masses $M_i$ ($i=1,2$) for each of
the species are
\begin{equation}
M_{i} = -(-1)^i M-{\frac{\Lambda}{2 \pi \theta}} \left({\sqrt{
1+{\frac{M^2}{\Lambda^2}} }}-{\frac{M}{\Lambda}}\right)
\label{eq:m}
\end{equation}
where $M$ is the solution of the gap equation (\ref{eq:newgap}) and
it is a function of the coupling constant ${\bar g}$. Hence, the
effective masses $M_i$ are also functions of the coupling constant. In
Fig.
{}~\ref{fig1} we show the qualitative form of the functions $M_i({\bar
g})$ for the entire range of couplings. Given the relation between
${\bar g}$ and the anisotropy $\lambda$, the curves of
Fig.
{}~\ref{fig1} will
help us to determine the phase diagram.

Thus,
while the spectrum of the semiclassical theory consists of two  massless
fermions for all ${\bar g} \leq {\bar g}_c$ and two massive fermions
(but whose masses have opposite signs) for ${\bar g} > {\bar g}_c$, the
quantum fluctuations of the gauge fields make the fermion spectrum
generically massive for {\em all} values of ${\bar g}$ ( and, hence for
all $\lambda$). However, eq.~(\ref{eq:m}) shows that, for $\theta$
positive and ${\bar g} \leq {\bar g}_c$, $M_1=M_2=-{\frac{\Lambda}{2 \pi
\theta}}<0$. In fact, for $\theta>0$, $M_2$ is always negative. However,
$M_1$ goes through zero and changes sign at some critical value of the
coupling constant ${\bar g}^* (\theta)$
\begin{equation}
{\bar g}^*(\theta)={\bar g}_c {\sqrt{1+{\frac{1}{\pi \theta}}}}
\label{eq:g*}
\end{equation}
For the value of $\theta=1/2 \pi$, of interest for the Heisenberg
antiferromagnet, we get ${\bar g}^*={\sqrt{3}} {\bar g}_c$.

Thus, we
conclude that the quantum fluctuations yield the following spectrum for
the fermions. For ${\bar g}\leq {\bar g}^*$ the two species of fermions
have masses which, for  general values of ${\bar g}$ have different
absolute values but have the {\em same} sign. This sign is {\em
opposite} to the sign of $\theta$. In contrast, for ${\bar g} > {\bar
g}^*$ the two fermionic species have masses with different absolute
values {\em and} opposite signs. Precisely {\em at} ${\bar g}^*$, the
mass of one of the species of fermions passes through zero. We should
regard this phenomenon as a {\em phase transition} and ${\bar g}^*$ as a
critical point. From the arguments presented above we should expect that
the critical value ${\bar g}^*$ should correspond to a critical
value of $\lambda$, which we will denote by $\lambda^*$, and that there
should be a phase transition at $\lambda^*$ in the anisotropic quantum
Heisenberg antiferromagnet~\cite{matthew}.

Given that the fluctuations make such important effects already at the
level of the leading corrections to the AFA, it is natural to inquire
what are the effects of even higher order corrections.
It is clear that nothing special happens at ${\bar g}_c$ and that it
does not correspond to a phase transition which has been shifted to
${\bar g}^*$. The apparent discontinuity in the derivative of the masses
at ${\bar g}_c$ is an artifact of the leading order calculation. Higher
order terms will smooth out this spurious effect. The actual value of
the critical coupling will also be renormalized by higher order terms.

We now will argue that the phase transition at ${\bar g}={\bar g}^*$
should
be identified with the {\em isotropic} Heisenberg antiferromagnet. The
argument in support of this identification relies on the counting of
massless excitations of the system. In turn, we will identify
these massless excitations with the Goldstone modes of the
antiferromagnet.

At this level of approximation, the spectrum of fermions
is massive for generic
values of the anisotropy with one of the species becoming massless just
at the critical coupling. Let us now investigate the spectrum of the
bosonic excitations, $A_\mu$ and $A_0^{\pi \pi}$, as a function of ${\bar
g}$. Since the fermions are massive, it is possible to integrate them
out off the partition function and to find an effective action for the
Bose fields which is local at length scales long compared with the
inverse of the mass. From the work of Deser, Jackiw and
Templeton~\cite{jackiw} we know that the long distance effective
Lagrangian for the
gauge field, ${\cal L}_{\rm ind} (A_\mu)$, induced by the fluctuations
of a fermion of mass $M$, is of the form
\begin{equation}
{\cal L}_{\rm ind} (A_\mu) \approx -{\frac{1}{4 \kappa^2}} F^{\mu \nu}
F_{\mu \nu}+{\frac{{\rm sign}(M)}{4 \pi}} \epsilon_{\mu \nu \lambda}A^\mu
F^{\nu \lambda}
\label{eq:Lind}
\end{equation}
where we recognize the last term as a Chern-Simons term and the sign of
its coupling constant is {\em equal} to the sign of the fermion mass
$M$. The parameter  $\kappa$ is proportional to $\vert M \vert$. A
similar analysis implies that the fluctuations of $A_0^{\pi \pi}$
are always massive.
However, since the fermions are actually massive at ${\bar
g}_c$, the fluctuations of the Ne{\` e}l order parameter field $A_0^{\pi
\pi}$ {\em never} become critical.

The total effective action for the fluctuating gauge field is the sum of
three contributions: (a) the (bare) Chern-Simons term with coupling
constant $\theta$, and (b) two additional Chern-Simons terms each with
the value of eq.~(\ref{eq:Lind}). Thus, the {\em total} Chern-Simons
coupling constant, $\theta_{\rm eff}$ is
\begin{equation}
\theta_{\rm eff}({\bar g})= \theta+{\frac{{\rm sign}(M_1({\bar g})
+M_2({\bar g}))}{4 \pi}}
\label{eq:theta}
\end{equation}
We find,
\begin{equation}
\theta_{\rm eff}({\bar g})= \left\{
\begin{array}{ll}
\theta-{\frac{2}{4 \pi}} & \mbox{if ${\bar g} \leq {\bar g}^*$}\\
\theta & \mbox{otherwise}
\end{array}
\right.
\label{eq:lenz}
\end{equation}
Thus, in the phase in which the effective fermion masses have the {\em
same} sign, the fluctuations of the fermions act so as to reduce (or
{\em screen}) the bare value of the Chern-Simons coupling constant
$\theta$. This is the ``Lenz law of statistics" referred to in the
Introduction.

For the particular case of interest for the antiferromagnet,
$\theta=1/2 \pi$, the screening is {\em  complete} and $\theta_{\rm
eff}=0$ for the case of bosons!. Thus, the Chern-Simons term is
cancelled out from the effective action of the gauge field which now
represents massless excitations. This cancelation of the Chern-Simons
term from the effective action of the gauge field is the anyon
superfluid scenario of references~\cite{laughlin,cwwh,anyons}. In
the anyon superfluid, the transverse massless gauge field  is
interpreted as the Goldstone boson of the superfluid state. We identify
this regime with the $XY$ phase of the antiferromagnet.

Conversely, for ${\bar g} > {\bar g}^*$, we get $\theta_{\rm
eff}=\theta$ and all collective modes are massive. In this phase the
Ne{\` e}l order parameter is non zero and {\em all} excitations have a
gap. This is the {\em Ising} phase of the antiferromagnet.

The phase diagram is now {\em almost} complete. Two issues still need to be
resolved: (1) the nature of the transition point at ${\bar g}^*$ and (2)
are the fermion states really part of the spectrum for any value of the
coupling constant?

These two problems actually are not independent from each other. Let us
first consider
the fate of the fermions.  For ${\bar g} > {\bar g}^*$, the fermions
have non-zero masses with opposite sign. In this phase, the effective
Chern-Simons coupling constant is equal to $1/2 \pi$. Hence, in this
phase, by the standard argument of statistical transmutation,
Chern-Simons gauge field turns the massive {\em fermions} into massive
{\em bosons}. However, for ${\bar g}< {\bar g}^*$,
the long distance effective action
for the gauge field does not include a Chern-Simons, which cancels out
but, instead,
the leading term has a Maxwell form. Thus, the actual physical mass (or
gap) of the fermion will be significantly renormalized by the
fluctuations of the transverse massless collective mode (the gauge
field). The result, exactly as in the case of the semion
superfluid~\cite{laughlin}, is that due to the quantum fluctuations of
the collective mode, the fermion acquires a logarithmically divergent
mass and, therefore, it disappears from the physical spectrum.

The presence of infrared divergent corrections to the fermion
self-energy in the range ${\bar g}< {\bar g}^*$ has
important consequences for the Ne{\` e}l order.
Eq.~(\ref{eq:contspe}) relates the expectation value of the
Ne{\` e}l order parameter ${\bar \Psi} T^3 \Psi$ to the  expectation
value of the field $A_0^{\pi \pi}$ (this equation is valid beyond the
saddle point approximation
provided that the field $A_0^{\pi \pi}$ is replaced by its exact
expectation value). At the level of the saddle-point, the expectation
value of $A_0^{\pi \pi}$ is equal to $M {\bar g}$. Higher order
corrections involve fermion self-energy insertions in the low order
diagrams. For ${\bar g}>{\bar g}^*$, these corrections are finite in
the infrared. However, for ${\bar g}<{\bar g}^*$, these
corrections are infrared divergent (due to the fermion self-energy
insertions) and negative (since
they have to give a value smaller than the classical result). The
precise computation of these effects to all orders is difficult and
beyond the domain of perturbation theory.
The presence of these infrared divergent contributions, already in the
leading corrections,
suggests that the Ne{\` e}l order is unstable for ${\bar g}<{\bar
g}^*$ and that the exact
expectation value of this operator has to vanish in that regime.
An alternative
picture of this effect can be seen by noticing that, as the coupling
constant $\bar g$ is decreased from large values ( that  is from the
classical Ne{\` e}l regime) and one of the masses becomes small,
tunneling processes between the two Ne{\` e}l states become increasingly
favorable. In particular,
the magnitude of the energy per
unit length of a Ne{\` e}l {\em domain wall} is set by the mass of the
fermionic excitations. Thus, in the regime in which the fermionic
excitations have a finite mass, the energy per unit length of the domain
wall is finite. Below ${\bar g}^*$, the
infrared divergencies in the fermion self-energy will force the energy per
unit length of the wall to vanish. Consequently,
the domain walls
will condense in this regime and will destroy the Ne{\` e}l long range
order. Precisely at ${\bar g}^*$, the energy per unit length of the
domain wall is still finite since there is still Ne{\` e}l order.
Thus, the domain wall energy should drop
to zero with a jump at ${\bar g}^*$.
The domain wall condensation as a mechanism for
the
destruction of long range order is well known in $1+1$-dimensional systems,
where the solitons play the role of the domain wall.
Hence, we conclude that the Ne{\` e}l order parameter should drop
to zero discontinuously at ${\bar g}^*$ and to vanish for all ${\bar
g}<{\bar g}^*$.

In
contrast, the operator which creates (or removing) one
fermion {\em and} one flux quantum simultaneously, is gauge invariant
and it has a {\em finite} mass. In anyon superfluidity this state is
usually called the vortex and it has the statistics of an anyon. In our
problem, again by statistical transmutation, it is a massive {\em
boson}. The mass of this state should scale  with  $M_1$. Hence, it
should vanish {\em exactly} at ${\bar g}^*$. In the lattice
Chern-Simons theory this state is created by an operator which changes
both charge and flux. For the case of the antiferromagnet, the operator
which creates this state is $S^+$. Thus, we argue that the massless
fermion of our spectrum is actually the extra Goldstone boson of the
antiferromagnet. Since this state becomes massless only at ${\bar g}^*$,
we identify this phase transition with the {\em isotropic} quantum
antiferromagnet. This point is then viewed as a limiting point and it
has the attributes of both phases. In particular, it has a non-vanishing
value for the Ne{\` e}l order parameter and two Goldstone
bosons~\cite{commentlambda}. We also note, in passing, that in
one-dimension a strikingly similar picture of the nature of the
isotropic point (from
the point of view of the symmetry analysis) was developed by Luther and
Peschel~\cite{luther}.

However, in the case of our problem, unlike the case of the
one-dimensional spin chain, there are no
non-perturbative methods available, such as bosonization, that will
allow for an exact treatment of the long distance behavior. The
topological invariance of the Chern-Simons theory strongly suggests that
no further renormalizations occur and that we have successfully
characterized the infrared stable fixed points. A detailed analysis of
the phase transition at ${\bar g}^*$, particularly the determination of
its universality class, critical exponents, {\it etc.}, requires a more
sophisticated
analysis of the effective theory than the one performed here.
The phase transition at ${\bar g}_c$ is in the universality class of
theories of self-interacting relativistic fermions (of the Gross-Neveu
type). Our analysis shows that this fixed point is unstable and does not
represent the long distance behavior of the system.
It is interesting to note that the removal of the phase transition at
${\bar g}_c$ and
its replacement by the transition at ${\bar g}^*$ where the Ne{\` e}l
order parameter drops to zero {\em discontinuously}, is strongly
reminiscent
of the physics of fluctuation induced first order transitions.

We can also give a renormalization group picture for the
arguments presented above. The AFA or, equivalently, the semiclassical theory
of this section, does not represent faithfully all the relevant
fluctuations
of the system. In particular, the {\em infrared stable} fixed point associated
with the $XY$ phase is simply not present in the semiclassical theory. The
quantum fluctuations of the gauge field contain the appropriate relevant
operator, the parity breaking induced mass term. Once this operator is
generated, the flow of the coupling constants is drastically changed as we
explore the low energy regime. In particular, the effective action of the
gauge fields acquires a finite renormalization of its Chern-Simons coupling
constant which tends to screen the statistics. The fermions are also affected
by this flow since the physical low energy mass (or energy gap) of the
spectrum is altered by the fluctuations of the bosons. In the $XY$ phase, the
excitations of the gauge fields are massless and their quantum
fluctuations suppress the fermions
from the spectrum. In the Ising phase, their fluctuations turn the fermions
into bosons. Clearly at both  stable fixed points, parity {\em and} time
reversal are not broken (for $\theta=1/2 \pi$).

We conclude this section with a few remarks. Firstly, our arguments show
that, for all values of the anisotropy, there are no states in the
spectrum with the quantum numbers of a fermion. All the states are
bosonic. This is not an accident since the system is not in a spin
liquid state for all values of the anisotropy. Nevertheless, given the
analogies between the AFA and the flux phases of the
theories of frustrated antiferromagnets, our results should be viewed as
an indication that, once fluctuations are fully taken into account, the
flux phases could become more like the standard phases of
antiferromagnets. We have presented qualitative arguments which show
that the phase transition at ${\bar g}^*$ can be viewed, in some sense, as a
first order transition since the order parameter must have a jump at that
point. A more rigurous proof of this statement still needs to be constructed.
Nevertheless, the arguments presented above show clearly the physical
mechanism behind this phenomenon.
A direct computation of the spin correlation
function $\langle S^{+} (x) S^{-} (x') \rangle$ should demonstrate quite
explicitly the presence of an additional massless state at the critical
point. Work on this problem is currently in progress.

\section{conclusions}
\label{sec-conclusions}

In this paper we have presented a field theory for the anisotropic
quantum
antiferromagnet on a square lattice, based on the Chern-Simons or
Wigner-Jordan approach. We discussed in detail the phase diagram, as a function
of anisotropy, at the level of the Average Field Approximation. We found
a continuous, second order phase transition at a critical anisotropy from a
flux-like phase to an Ising phase. We showed that this phase transition is
spurious and that the massless flux phase cannot possibly describe the the $XY$
regime of the quantum antiferromagnet. We used a semiclassical theory,
based in the method of spin coherent
states, to derive an anisotropic non-linear sigma model for this system
and derived a phase diagram for the system, valid in the limit of large
spin. We considered the role of fluctuations around the AFA and
showed that they induce relevant operators, not included in the AFA,
which drive the low energy behavior of the system. We
derived an effective field theory of self interacting fermions coupled to
Chern-Simons gauge fields and showed that its fluctuations contain all the
necessary relevant operators to yield a correct phase diagram. In
particular they induce $P$ and $T$ symmetry breaking
fermion mass terms which should necessarily be present for arbitrary
values of the Chern-Simons coupling constant.
We gave a set of arguments which indicate that the fluctuations of the
gauge field drive the theory away from the universality class of
theories of self-interacting relativistic fermions (of the Gross-Neveu
type). We identified the infrared stable fixed points and showed that
the spectrum of the system at theses fixed points coincides with the
expected spectrum of the phases of the antiferromagnet discussed in
section \ref{sec-1/s}.

We conclude with a comment on the accuracy of the AFA. The AFA has become a
standard tool and it is widely used in a variety of fields of condensed
matter,
most prominently in theories of the Fractional Quantum Hall Effect. The
difficulties that we found in applying the AFA to the quantum Heisenberg
antiferromagnet show that this approximation can be unreliable if the
resulting fermion spectrum is massless. In such situation fluctuations may
(and generally will) induce relevant operators which will necessarily generate
gaps in the fermionic spectrum. This is not a problem for theories of the
incompressible states of the FQHE but could well be the case for
the compressible even denominator states.

\section{Acknowledgements}

A.~G.~R.~ thanks the Department of Physics of the University of
Illinois for its kind hospitality. This work was done in part while
A.~G.~R.~ was a member of the Department of Physics of the
University of Tennessee at Knoxville. This work was supported in part by
the
National Science Foundation through the grants NSF DMR-9015771 at the
University of Tennessee and NSF STC-9120000 at the University of Chicago
(AGR), NSF DMR-91-22385 at the University of Illinois at
Urbana-Champaign (EF),
by a Fellowship of the American Association of University Women
(AL), and
by the U.S. Department of Energy through contract DE-AC05-84OR21400
administered by Martin Marietta Energy Systems Inc.


\begin{figure}
\caption[]{
Fermion mass gaps. The full curves show the mass of
the fermion species
$m_i=M_i/\Lambda$ ($i=1,2$)
against the (normalized) coupling constant
$t=\bar g / \bar g_c$.
The broken curves are the fermion masses
$m=M/\Lambda$,
predicted by the AFA. The phase transition occurs at
$t^* = {\sqrt{3}}$
where the mass $m_1$ vanishes. The range of the
Ising and $XY$ phases is also shown.}
\label{fig1}
\end{figure}


\newpage

\begin{references}

\bibitem{bednorz} J.~A.~Bednorz and K.~A.~Muller, {\sl Z.~Phys.~\/}
{\bf B64}, 189 (1986).
\bibitem{anderson} P.~W.~Anderson, {\sl Science} {\bf 235}, 1196 (1987).
\bibitem{bethe} H.~Bethe {\sl Z.~Phys.\/ } {\bf 71}, 205 (1931)
\bibitem{jordan} P.~Jordan and E.~P.~Wigner {\sl Z.~Phys.\/ } {\bf 47},
631 (1928).
\bibitem{liebmattis} E.~Lieb, T.~Schultz and D.~C.~Mattis, {\sl
Ann.~Phys.~(N.~Y.~)\/} {\bf 16\/}, 407 (1961).
\bibitem{holstein} T.~Holstein and H.~Primakoff, {\sl Phys.~Rev.~} {\bf
58\/}, 1098 (1940).
\bibitem{kittel} C.~Kittel, {\sl Quantum Theory of Solids}, John Wiley
\& Sons , 1987 Chapter 4.
\bibitem{haldanegap} F.~D.~M.~Haldane, {\sl Phys.~Lett.~\/} {\bf A93\/},
464 (1983).
\bibitem{luther} A.~Luther and I.~Peschel, {\sl Phys.~Rev.~\/} {\bf
B12\/}, 3908 (1975).
\bibitem{duncan} F.~D.~M.~Haldane, {\sl Phys.~Rev.~Lett.~\/}
{\bf 50\/}, 1153 (1983).
\bibitem{spinwave} E.~Manousakis, {\sl Rev.~Mod.~Phys.~\/} {\bf 63\/}, 1
(1991).
\bibitem{xy} T.~Kennedy, E.~H.~Lieb and B.~S.~Shastry, {\sl
Phys.~Rev.~Lett.~\/}
{\bf 61}, 2582 (1988); K.~Kubo, and T.~Kishi, {\sl Phys.~Rev.~Lett.~\/}
{\bf 61}, 2585 (1988).
\bibitem{dyson} F.~Dyson, E.~H.~Lieb and B.~Simon, {\sl
J.~Stat.~Phys.~\/} {\bf 18\/}, 335 (1978).
\bibitem{reger} J.~D.~Reger and A.~P.~Young, {\sl Phys.~Rev.~\/} {\bf
B37\/}, 5024 (1988).
\bibitem{manousakis} E.~Manousakis, reference\cite{spinwave}, and
references therein.
\bibitem{vmc} S.~Liang, B.~Doucot and
P.~W.~Anderson,
{\sl Phys.~Rev.~Lett\/} {\bf 61\/}, 365 (1988); S.~Liang, {\sl
Phys.~Rev.~Lett.~\/} {\bf 64\/}, 1597 (1990).
\bibitem{1/s} T.~Dombre and N.~Read, {\sl Phys.~Rev.~\/} {\bf B38\/},
7181 (1988); E.~Fradkin and M.~Stone {\sl Phys.~Rev.~\/} {\bf B38\/},
7215(1988); F.~D.~M.~Haldane, {\sl Phys.~Rev.~Lett.~\/} {\bf 61\/}, 1029
(1988); L.~Ioffe and A.~Larkin, {\sl Int.~J.~Mod.~Phys.~\/} {\bf B2},
203 (1988); X.~G.~Wen and A.~Zee, {\sl Phys.~Rev.~Lett.~\/} {\bf 61\/},
1025 (1988).
\bibitem{chn} S.~Chakravarty, B.~I.~Halperin and D.~Nelson,
{\sl Phys.~Rev.~Lett\/} {\bf 60\/}, 1057 (1988).
\bibitem{fradkin1} E.~Fradkin, {\sl Phys.~Rev.~Lett\/} {\bf 63\/}, 322
(1989).
\bibitem{laughlin} R.~B.~Laughlin, {\sl Phys.~Rev.~Lett\/} {\bf 60\/},
2677 (1988); C.~B.~Hanna,
R.~B.~Laughlin and A.~L.~Fetter, {\sl Phys.~Rev.~\/} {\bf B40\/}, 8745
(1989).
\bibitem{cwwh} Y.~-H.~Chen, F.~Wilczek, E.~Witten, and B.~I.~Halperin,
{\sl Int.~J.~Mod.~Phys.~\/B} {\bf B3}, 1001 (1989).
\bibitem{anyons} E.~Fradkin, {\sl Phys.~Rev.~\/} {\bf B42\/}, 570
(1990).
\bibitem{zhk} S.~C.~Zhang, T.~Hansson and S.~Kivelson,
{\sl Phys.~Rev.~Lett\/} {\bf 62\/}, 82 (1989).
\bibitem{lopez} Ana Lopez and  Eduardo Fradkin, {\sl Phys.~Rev.~\/}
{\bf B44\/}, 5246, (1991).
\bibitem{pwa} P.~W.~Anderson, {\sl Science\/} {\bf 235}, 1196 (1986).
\bibitem{affleck}  I.~K.~Affleck and J.~B.~Marston, {\sl Phys.~Rev.~\/}
{\bf B37\/}, 3774 (1988).
\bibitem{kotliar} G.~B.~Kotliar,  {\sl Phys.~Rev.~\/}  {\bf B37\/},
3664 (1988).
\bibitem{wang} Y.~R.~Wang, {\sl Phys.~Rev.~\/} {\bf B43\/}, 3786
(1991).
\bibitem{jackiw} S.~Deser, R.~Jackiw and S.~Templeton,
{\sl Phys.~Rev.~Lett\/} {\bf 48\/}, 372 (1982).
\bibitem{steve} We thank S.~Kivelson for suggesting this interpretation.
\bibitem{matthew} This mechanism is somewhat analogous to the
metal insulator transition recently studied by W.~Chen, M.~P.~A.~Fisher
and Y.~S.~Wu (CFW), {\sl Phys.~Rev.~\/} {\bf B 48 \/}, 13749 (1993).
Unlike the problem studied by CFW, in the mechanism that we use here the
fermionic masses are actual fluctuating fields which, for some range of
$\lambda$, acquire non-vanishing expectation values.
\bibitem{hlm} B.~I.~Halperin, T.~Lubensky and S.~K.~Ma,
{\sl Phys.~Rev.~Lett\/} {\bf 32\/}, 292 (1974).
\bibitem{cw} S.~Coleman and E.~Weinberg, {\sl Phys.~Rev.~\/} {\bf B7\/},
1888 (1973).
\bibitem{fradkin2} Eduardo Fradkin, {\sl Field Theories of Condensed Matter
Systems}, Addison \& Wesley, 1991.
\bibitem{wilczek} F.~Wilczek, {\sl Phys.~Rev.~Lett\/} {\bf
49\/}, 957 (1982), and {\sl Phys.~Rev.~Lett\/} {\bf 48\/}, 1144 (1982).
\bibitem{eliezer} D.~Eliezer and G.~W.~Semenoff, {\sl Phys.~Lett\/} {\bf
B286\/}, 118 (1992), and references therein; this paper corrects the
commutation relations of reference \cite{fradkin1} which
are inconsistent.
\bibitem{girvin}
G.~S.~Canright, S.~M.~ Girvin, and A.~Brass,
{\sl Phys.~Rev.~Lett\/} {\bf 63\/}, 2295 (1989).
\bibitem{resfqhe} Ana Lopez and  Eduardo Fradkin, {\sl Phys.~Rev.~\/}
{\bf B47\/}, 7080 (1993).
\bibitem{hofstadter} D.~Hofstadter, {\sl Phys.~Rev.~\/}{\bf B14\/}, 2239
(1976).
\bibitem{huse} D.~A.~Huse and V.~Elser, {\sl Phys.~Rev.~Lett\/}
{\bf60\/}, 2531 (1988).
\bibitem{emery} A detailed derivation of this equivalence can be found
in V.~J.~Emery, in {\sl Highly Conducting One-Dimensional Solids\/},
edited by J.~Devreese, R.~Evrard and V.~Van Doren (Plenum, New York,
1979).
\bibitem{lieb} E.~H.~Lieb and D.~C.~Mattis, {\sl Mathematical Physics in
One Dimension\/}, ( Academic Press, New York, 1966).
\bibitem{wwz} X.~G.~Wen, F.~Wilczek and A.~Zee,  {\sl
Phys.~Rev.~\/}{\bf B39\/}, 11413 (1989).
\bibitem{gn} In this form, this theory is known as the one-component
Gross-Neveu model, D.~J.~Gross and A.~Neveu, {\sl Phys.~Rev.~\/}{\bf
D10\/}, 3235 (1974).
We are following here reference\cite{fradkin2}.
\bibitem{maxwell} The last term in eq.~(\ref{eq:Lint}) is just equal
to ${\frac {1}{2 {\bar e}^2}} B^2$ and it is not a full Maxwell
Lagrangian which, instead, should look like ${\frac {1}{2 {\bar e}^2}}
({\frac {{\bf E}^2}{v_{\rm G}^2}}-B^2)$ where $v_{\rm G}$ is the speed
of propagation of the gauge fields. Such electric-like terms are absent
at the present order but do get generated by renormalization effects. In
any event, both the coupling constant (or effective ``charge") $\bar e$
and the speed $v_{\rm G}$ are dimensionful parameters. A full
Maxwell-type term is still irrelevant at long distances. Thus,  in
principle, such terms can (and will) be ignored.
\bibitem{sign} We use the statistical mechanical sign in the
definition of the $\beta$-function.
\bibitem{dennijs} M.~P.~M.~den Nijs, {\sl Phys.~Rev.~\/}{\bf B23\/},
6111 (1981).
\bibitem{umpklapp}F.~D.~M.~Haldane, {\sl Phys.~Rev.~\/}{\bf B25\/}, 4925
(1982).
\bibitem{weichen} Wei Chen and Miao Li, {\sl Phys.~Rev.~Lett.\/}{\bf
70\/}, 884 (1993).
\bibitem{commentlambda} The critical value ${\bar g}^*$ corresponds to a
$\lambda^* \approx {\sqrt{3}}/4 <1$. The fact that it is not precisely
equal to one (the nominal value of the anisotropic point) is due to the
approximations that went in taking the continuum limit. The same happens
in the bosonization approaches (based in the Wigner-Jordan
transformation) in one space dimension ~\cite{luther}. In any case the
symmetries and, consequently, the universality class, are the same.

\end{references}
\end{document}